\documentclass[useAMS,usenatbib,usegraphicx,fleqn]{mn2e}

\usepackage[T1]{fontenc}
\usepackage{pslatex}
\usepackage{color}
\usepackage{amsmath}
\usepackage{url}

\def\gtrsim{\mathrel{\hbox{\rlap{\hbox{\lower3pt\hbox{$\sim$}}}\hbox{\raise2pt\hbox{$>$}}}}}

\definecolor{orange}{rgb}{1,0.3,0}
\definecolor{purple}{rgb}{1,0,1}

\newcommand{\chone}{3.6\,\micron~}
\newcommand{\chtwo}{4.5\,\micron~}

\title[Stellar content of SZ clusters]{The Atacama Cosmology Telescope: the stellar content of 
galaxy clusters selected using the Sunyaev-Zel'dovich effect}
\author[Hilton et al.]
{\parbox{\textwidth}{\raggedright Matt~Hilton,$^{1,2}$\thanks{E-mail: hiltonm@ukzn.ac.za}
Matthew~Hasselfield,$^{3}$
Crist\'{o}bal~Sif\'{o}n,$^{4}$
Andrew~J.~Baker,$^{5}$
L.~Felipe~Barrientos,$^{6}$
Nicholas~Battaglia,$^{7,8}$
J.~Richard~Bond,$^{8}$
Devin~Crichton,$^{9}$
Sudeep~Das,$^{10,11}$
Mark~J.~Devlin,$^{12}$
Megan~Gralla,$^{9}$
Amir~Hajian,$^{8}$
Adam~D.~Hincks,$^{8}$
John~P.~Hughes,$^{5}$
Leopoldo~Infante,$^{6}$
Kent~D.~Irwin,$^{13}$
Arthur~Kosowsky,$^{14}$
Yen-Ting~Lin,$^{15}$
Tobias~A.~Marriage,$^{9}$
Danica~Marsden,$^{16}$
Felipe~Menanteau,$^{5}$
Kavilan~Moodley,$^{2}$
Michael~D.~Niemack,$^{17}$
Mike~R.~Nolta,$^{8}$
Lyman~A.~Page,$^{18}$
Erik~D.~Reese,$^{12}$
Jon~Sievers,$^{18,8}$
David~N.~Spergel$^{19}$ and
Edward~J.~Wollack$^{20}$}\vspace{0.4cm}\\
\parbox{\textwidth}{\raggedright $^{1}$~Centre for Astronomy \& Particle Theory, School of Physics and Astronomy, University of Nottingham, NG7 2RD, UK\\
$^{2}$~Astrophysics \& Cosmology Research Unit, School of Mathematics, Statistics \& Computer Science, University of KwaZulu-Natal, Durban 4041, SA\\
$^{3}$~Department of Physics and Astronomy, University of British Columbia, Vancouver, BC, Canada V6T 1Z4\\
$^{4}$~Leiden Observatory, Leiden University, PO Box 9513, NL-2300 RA Leiden, Netherlands\\
$^{5}$~Department of Physics and Astronomy, Rutgers, The State University of New Jersey, 136 Frelinghuysen Road, Piscataway, NJ USA 08854-8019\\
$^{6}$~Departamento de Astronom{\'{i}}a y Astrof{\'{i}}sica, Facultad de F{\'{i}}sica, Pontific\'{i}a Universidad Cat\'{o}lica, Casilla 306, Santiago 22, Chile\\
$^{7}$~Department of Physics, Carnegie Mellon University, Pittsburgh, PA 15213\\
$^{8}$~Canadian Institute for Theoretical Astrophysics, University of Toronto, Toronto, ON, Canada M5S 3H8\\
$^{9}$~Department of Physics and Astronomy, The Johns Hopkins University, 3400 N. Charles St., Baltimore, MD 21218-2686\\
$^{10}$~Argonne National Laboratory, 9700 S. Cass Ave., Lemont,
IL USA 60439\\
$^{11}$~Berkeley Center for Cosmological Physics, LBL and Department of Physics, University of California, Berkeley, CA, USA
94720\\
$^{12}$~Department of Physics and Astronomy, University of Pennsylvania, 209 South 33rd Street, Philadelphia, PA, USA 19104\\
$^{13}$~NIST Quantum Devices Group, 325 Broadway Mailcode 817.03, Boulder, CO, USA 80305\\
$^{14}$~Department of Physics and Astronomy, University of Pittsburgh, Pittsburgh, PA, USA 15260\\
$^{15}$~Institute of Astronomy and Astrophysics, Academia Sinica, Taipei, Taiwan\\
$^{16}$~Department of Physics, University of California Santa Barbara, CA USA 93106\\
$^{17}$~Department of Physics, Cornell University, Ithaca, NY, USA 14853\\
$^{18}$~Joseph Henry Laboratories of Physics, Jadwin Hall, Princeton University, Princeton, NJ, USA 08544\\
$^{19}$~Department of Astrophysical Sciences, Peyton Hall, Princeton University, Princeton, NJ, USA 08544\\
$^{20}$~NASA/Goddard Space Flight Center, Greenbelt, MD, USA 20771\\}
}

\begin{document}

\renewcommand{\bottomfraction}{0.95}

\date{Draft version: \today}

\pagerange{\pageref{firstpage}--\pageref{lastpage}} \pubyear{2013}

\maketitle

\label{firstpage}

\begin{abstract}
We present a first measurement of the stellar mass component of galaxy clusters selected via the
Sunyaev-Zel'dovich (SZ) effect, using \chone and \chtwo photometry from the \textit{Spitzer Space Telescope}. 
Our sample consists of 14 clusters detected by the Atacama Cosmology Telescope
(ACT), which span the redshift range $0.27 < z < 1.07$ (median $z = 0.50$), and have dynamical 
mass measurements, accurate to about 30 per cent, with median $M_{500} = 6.9 \times 10^{14}$\,M$_{\sun}$.
We measure the \chone and \chtwo galaxy luminosity functions, finding the characteristic magnitude ($m^*$)
and faint-end slope ($\alpha$) to be similar to those for IR-selected cluster samples. We perform the first 
measurements of the scaling of SZ-observables ($Y_{500}$ and $y_0$) with both brightest cluster galaxy (BCG) stellar
mass and total cluster stellar mass ($M_{500}^{\rm star}$). We find a significant correlation between BCG stellar
mass and $Y_{500}$ ($E(z)^{-2/3}\,D_A^2\,Y_{500} \propto M_*^{1.2 \pm 0.6}$), although we are not able to 
obtain a strong constraint on the slope of the relation due to the small sample size. Additionally, we obtain 
$E(z)^{-2/3}\,D_A^2\,Y_{500} \propto M_{500}^{\rm star}\,^{1.0 \pm 0.6}$
for the scaling with total stellar mass. The mass fraction in stars spans the range 0.006--0.034, with the 
second ranked cluster in terms of dynamical mass (ACT-CL~J0237-4939) having an unusually
low total stellar mass and the lowest stellar mass fraction. For the five clusters with gas mass measurements 
available in the literature, we see no evidence for a shortfall of baryons relative to the cosmic mean value.
\end{abstract}


\section{Introduction}
\label{s_intro}

Galaxy clusters are the most massive gravitationally bound objects in the Universe, and as such their changing 
abundance with redshift traces the process of structure formation. By studying the evolution of
their properties with redshift, we can learn the assembly history of their constituent dark matter, 
gas, and galaxies. The large masses of clusters ensure that the gas and stellar mass observed
within them remain gravitationally bound. They therefore represent a fair sample of the Universe
as it evolves over cosmic time. Measurements of cluster gas fractions have been used to constrain
cosmological parameters, including the dark energy equation of state \citep[e.g.,][]{Allen_2008}. While the hot
gas in clusters makes up the majority of the baryonic component (around 80 per cent), accurate measurements
of the mass fraction in the minority stellar component are also required to gain insight into the physical
processes occurring within clusters. 

It is known that the physics which determines the properties of the intracluster medium (ICM) is more 
complicated than simply the action of gravitational collapse alone, as the observed scaling relations, such as
between X-ray luminosity and temperature, deviate from the self-similar expectation 
\citep[e.g.,][]{Markevitch_1998, ArnaudEvrard_1999, Pratt_2009}, indicating that an additional source of
energy is heating the ICM. While some energy is injected by supernovae (SNe) within galaxies, it is likely that the
bulk of the energy comes from Active Galactic Nuclei (AGN) in the centres of clusters, as observations of low
redshift clusters show that AGN jets, seen in radio imaging, carve out cavities in the hot gas observed at 
X-ray wavelengths \citep[e.g.,][]{Birzan_2004, McNamara_2005, Blanton_2011}. Feedback by AGN and SNe must
also leave an imprint on the galaxy populations of clusters, and is a crucial component of galaxy formation
models \citep[e.g.,][]{DeLucia_2006, Bower_2006, Bower_2008}, where it is invoked to quench star formation in massive
haloes. Therefore, in principle, it is possible to constrain the strength of feedback processes by measuring the stellar 
fractions of clusters \citep*[e.g.,][]{Bode_2009}.

The baryon fractions of clusters in the local Universe have been measured by several studies \citep[e.g.,][]{Lin_2003,
Gonzalez_2007, Andreon_2010, Balogh_2011}. They find that the mass fraction contained in stars is smaller for more massive clusters, while conversely the gas 
fractions are larger \citep[e.g.,][]{Gonzalez_2007}. Within $R_{500}$ (the radius at which the enclosed density
is 500 times the critical density of the Universe), there may be a shortfall of baryons
with respect to the cosmic value inferred from measurements of the cosmic microwave background \citep[$f_{\rm b} = 0.166 \pm 0.020$;][]{Komatsu_2011},
with cluster studies finding $f_{\rm b} \approx 0.13$ \citep[e.g.,][]{Gonzalez_2007}, although the uncertainties
are large, and subject to various systematic effects. There are few measurements at intermediate redshift: \citet{Giodini_2009} 
describe measurements of the baryon fractions in $0.1 < z < 1$ groups and clusters found in the COSMOS \citep{Scoville_2007} 
field, while \citet{Lin_2012} present estimates for a heterogeneous sample of 45 clusters at $0.1 < z < 0.6$ using
archival X-ray data and infrared photometry from the Wide-Field Infrared Survey Explorer
\citep[WISE;][]{Wright_2010}. 

In this paper, we report a first characterisation of the stellar fractions, stellar mass scaling relations, and
properties of brightest cluster galaxies (BCGs), using infrared data from the \textit{Spitzer Space Telescope},
for a sample of galaxy clusters selected via the Sunyaev-Zel'dovich effect \citep[SZ;][]{SZ_1972} by the Atacama Cosmology Telescope
\citep[ACT;][]{Swetz_2011}. The SZ effect is the inverse Compton scattering of cosmic microwave background
photons by electrons in the hot gas atmospheres of clusters. In principle, it allows the construction of approximately mass-limited,
redshift-independent samples of clusters \citep*[see, e.g., the review by][]{Carlstrom_2002}. 

The structure of this paper is as follows. In Section~\ref{s_observations}, we briefly describe the 
cluster sample and associated ancillary data, which include dynamical mass estimates obtained from extensive spectroscopic
observations with 8\,m class telescopes \citep{Sifon_2012}. We also describe the processing of the \textit{Spitzer}
infrared imaging used in this work. In Section~\ref{s_LFs}, we present a measurement of the \chone and \chtwo galaxy
luminosity functions, and compare our findings with results obtained for IR-selected cluster samples. We examine
the IR-properties of the BCGs, and their scaling relations with dynamical mass and SZ observables, in 
Section~\ref{s_BCGs}. We similarly present scaling relations with total stellar mass, and the stellar fractions,
in Section~\ref{s_totalStellarMass}. We discuss our findings in Section~\ref{s_discuss}, and conclude
in Section~\ref{s_conclusions}.

\begin{figure*}
\includegraphics[width=18cm]{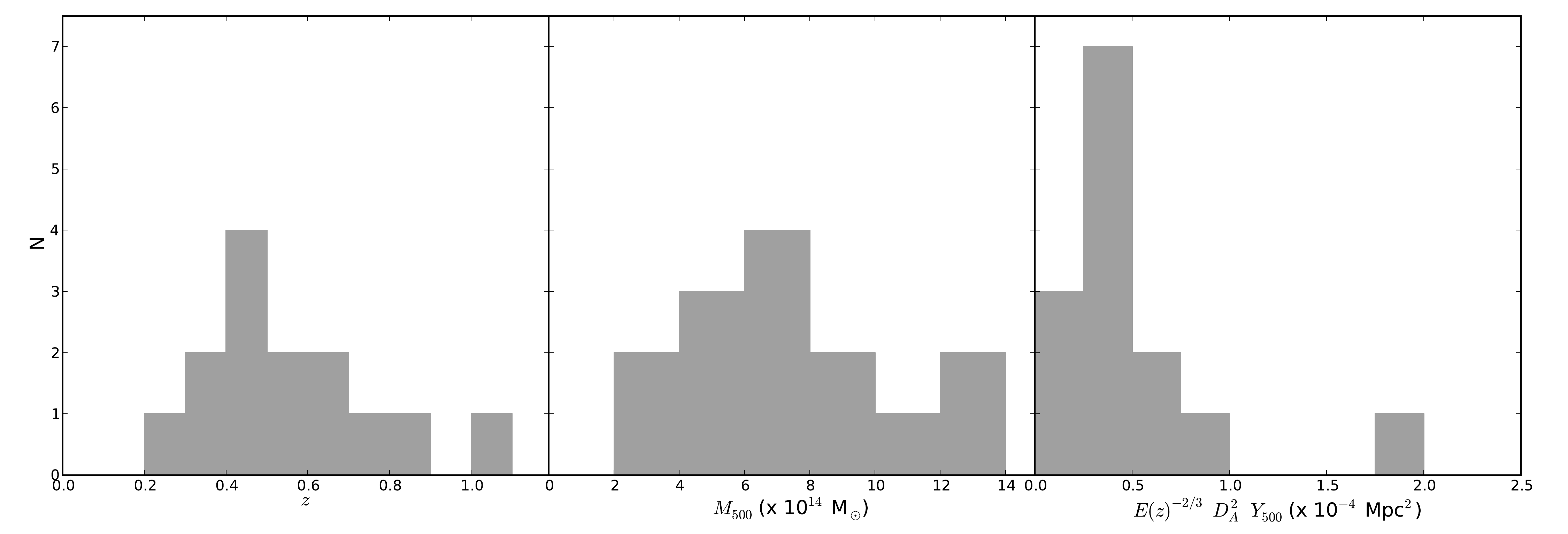}
\caption{Redshift, dynamical mass, and $Y_{500}$ distributions for the Sunyaev-Zel'dovich effect selected 
cluster sample used in this work \citep[see][for details]{Sifon_2012}. The clear outlier in the $Y_{500}$
distribution is the $z = 0.87$ merger system ACT-CL~J0102-4915 \citep[``El Gordo''; see][]{Menanteau_2012}.}
\label{f_sifonHistograms}
\end{figure*}

We assume a cosmology with $\Omega_{\rm m}=0.3$, $\Omega_\Lambda=0.7$, and
$H_0=70$~km~s$^{-1}$~Mpc$^{-1}$ throughout. All magnitudes are on the AB system \citep{Oke_1974}, unless
otherwise stated. We adopt a \citet{Salpeter_1955} Initial Mass Function (IMF) for stellar mass estimates.
All estimates of cluster masses and SZ-signals are measured within a characteristic radius defined 
with respect to the critical density at the cluster redshift.

\section{Data and Observations}
\label{s_observations}

\subsection{Cluster sample and optical observations}
The cluster sample used in this work is drawn from the \citet{Marriage_2011} SZ-selected cluster catalogue, 
detected in 148\,GHz observations (conducted in 2008) with ACT, a 6\,m telescope located in Northern Chile.
The pipeline used to produce maps (directly calibrated to WMAP; see \citealt{Hajian_2011}) from the ACT 
timestream data, is described in \citet{Dunner_2012}.
Follow-up optical imaging observations of these clusters, using the NTT and SOAR telescopes, are reported in 
\citet{Menanteau_2010}, and \citet{Sehgal_2011} discuss the cosmological constraints obtained from this 
sample.

\begin{table*}
\caption{Properties of the cluster sample used in this work. Masses are dynamical estimates
obtained from line of sight velocity dispersion measurements; all values have been rescaled 
from $R_{200}$ (as reported by \citealt{Sifon_2012}) to $R_{500}$.}
\label{t_clusterMasses}
\begin{tabular}{|c|c|c|c|c|c|}
\hline
Name & $z$ & $M_{500}$ ($10^{14}$\,M$_{\sun}$) & $Y_{500}$ ($10^{-11}$) & $y_0$ ($10^{-4}$) & $R_{500}$ (Mpc)\\
\hline
ACT-CL J0102-4915 & $0.870$ & \phantom{0}$9.8 \pm 2.3$ & \phantom{0}$9.7 \pm 1.7$ & $7.17 \pm 0.88$ & $1.1$\\
ACT-CL J0215-5212 & $0.480$ & \phantom{0}$5.8 \pm 1.7$ & \phantom{0}$2.0 \pm 0.8$ & $1.18 \pm 0.27$ & $1.1$\\
ACT-CL J0232-5257 & $0.556$ & \phantom{0}$4.2 \pm 1.5$ & \phantom{0}$1.0 \pm 0.6$ & $0.94 \pm 0.27$ & $0.9$\\
ACT-CL J0235-5121 & $0.278$ & \phantom{0}$7.0 \pm 2.1$ & \phantom{0}$5.0 \pm 1.7$ & $1.14 \pm 0.22$ & $1.2$\\
ACT-CL J0237-4939 & $0.334$ & $12.2 \pm 2.5$ & \phantom{0}$2.6 \pm 1.8$ & $1.09 \pm 0.31$ & $1.4$\\
ACT-CL J0304-4921 & $0.392$ & \phantom{0}$7.7 \pm 1.8$ & \phantom{0}$5.8 \pm 2.2$ & $2.08 \pm 0.41$ & $1.2$\\
ACT-CL J0330-5227 & $0.442$ & $10.3 \pm 2.4$ & \phantom{0}$4.6 \pm 1.0$ & $1.68 \pm 0.24$ & $1.3$\\
ACT-CL J0346-5438 & $0.530$ & \phantom{0}$6.4 \pm 1.4$ & \phantom{0}$2.8 \pm 0.9$ & $1.67 \pm 0.34$ & $1.1$\\
ACT-CL J0438-5419 & $0.421$ & $12.7 \pm 3.0$ & \phantom{0}$7.4 \pm 0.8$ & $2.07 \pm 0.17$ & $1.4$\\
ACT-CL J0509-5341 & $0.461$ & \phantom{0}$3.5 \pm 1.4$ & \phantom{0}$2.6 \pm 0.6$ & $1.13 \pm 0.19$ & $0.9$\\
ACT-CL J0528-5259 & $0.768$ & \phantom{0}$3.6 \pm 1.3$ & \phantom{0}$0.8 \pm 0.3$ & $1.03 \pm 0.28$ & $0.8$\\
ACT-CL J0546-5345 & $1.066$ & \phantom{0}$4.8 \pm 2.5$ & \phantom{0}$1.7 \pm 0.3$ & $2.54 \pm 0.39$ & $0.8$\\
ACT-CL J0559-5249 & $0.609$ & \phantom{0}$8.9 \pm 2.6$ & \phantom{0}$2.4 \pm 0.5$ & $1.48 \pm 0.23$ & $1.2$\\
ACT-CL J0616-5227 & $0.684$ & \phantom{0}$6.8 \pm 2.9$ & \phantom{0}$2.5 \pm 0.5$ & $1.87 \pm 0.28$ & $1.0$\\
\hline
\end{tabular}
\end{table*}

In this work, we use a subset of 14 clusters from the ACT sample that have received extensive spectroscopic 
observations with Gemini South and the VLT \citep{Sifon_2012}. On average, redshifts were 
measured for 60 galaxies per cluster, resulting in dynamical mass estimates with a typical uncertainty of 
about 30 per cent. These measurements were used to investigate the mass scaling relation with the SZ-signal 
characterised using three different methods. We use two of these types of measurements in this paper; both are
described in detail in \citet{Hasselfield_2012}. The first is the dimensionless, matched filter SZ amplitude 
($y_0$), which is extracted from maps filtered using an \citet{Arnaud_2010} profile with fixed angular scale 
$\theta_{500} = 5.9\arcmin$, and subsequently corrected to the angular scale corresponding to $R_{500}$ at
the observed cluster redshift as described in 
\citeauthor{Hasselfield_2012}. Our second estimate is the widely used, spherically integrated Compton 
signal ($Y_{500}$), measured within the radius $R_{500}$. It is important to note
that both of these measurements ultimately derive from the matched filter amplitude, and are therefore highly
correlated with each other and dependent upon the assumed \citet{Arnaud_2010} model for the cluster SZ signal. 
We use the updated $Y_{500}$ and $y_0$ values presented in the Appendix of \citeauthor{Hasselfield_2012}; these 
were measured using improved ACT maps which include data obtained in the 2009-2010 observing seasons. 

Note that because our IRAC data do not cover out to $R_{200}$ for the lower redshift objects in our 
sample (see Section~\ref{s_irac}), we rescale the measurements of \citet{Sifon_2012} at $R_{200}$ to 
$R_{500}$ using the concentration--mass relation of \citet[][]{Duffy_2008}. We include the uncertainty 
introduced by the scatter in this relation in the error bars on the $M_{500}$ measurements. This results in
the fractional mass errors being approximately 9 per cent larger than the uncertainties quoted in 
\citet{Sifon_2012}. Table~\ref{t_clusterMasses} lists the cluster properties.

Fig.~\ref{f_sifonHistograms} shows the redshift, mass, and $Y_{500}$ distributions for the clusters used in
this work. The sample spans the redshift range $0.27 < z < 1.07$, with median $z = 0.50$, and has median mass
$M_{500} = 6.9 \times 10^{14}$\,$M_{\sun}$. The object with the largest intrinsic $Y_{500}$, which is a clear
outlier from the distribution shown in the right panel of Fig.~\ref{f_sifonHistograms}, is the spectacular
``El Gordo'' (ACT-CL~J0102-4915) merger system at $z = 0.87$, for which we have previously published a 
detailed multiwavelength analysis \citep{Menanteau_2012}.

\subsection{\textit{Spitzer} IRAC imaging and photometry}
\label{s_irac}
\begin{figure}
\includegraphics[width=8.5cm]{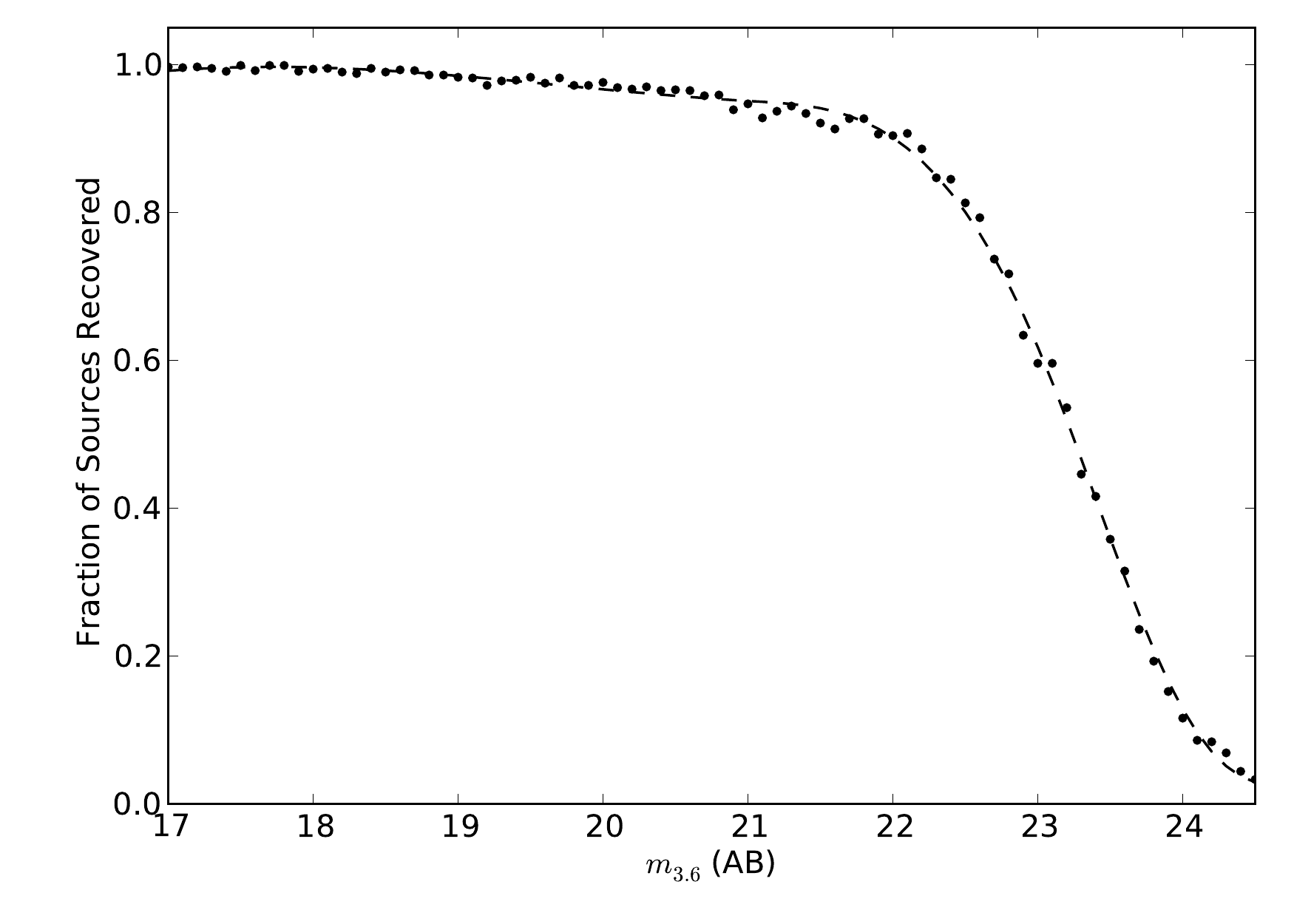}
\caption{Fraction of synthetic point sources recovered as a function of magnitude in the \chone channel. The
dashed line is a spline fit to the data. The 80 per cent completeness depth is approximately 22.6 mag.}
\label{f_completeness}
\end{figure}

We obtained IRAC \chone and \chtwo observations of the cluster sample during the period August--December 2010
(program ID: 70149, PI:Menanteau), using a $2 \times 2$ grid of IRAC pointings centred on each cluster position.
A total of $10 \times 100$ sec frames were obtained in each channel at each grid position, using a large scale
cycling dither pattern. The basic calibrated data (BCD) images were corrected for pulldown using the software
of Ashby \& Hora\footnotemark, and then mosaicked using MOPEX \citep{MakovozKhan_2005} to give images which are 
$\approx 13\arcmin$ on a side with 0.6$\arcsec$ pixel scale. The mosaics for each channel were then registered 
to a common pixel coordinate system. These mosaics cover out to $R_{500}$ for every cluster in the sample. 
By inserting synthetic point sources, we estimate the 80 per cent completeness depths of the final maps to be 
$\approx 22.6$ mag (AB) in both channels (see Fig.~\ref{f_completeness}).
\footnotetext{See \url{http://irsa.ipac.caltech.edu/data/SPITZER/docs/dataanalysistools/tools/contributed/irac/fixpulldown/}}

Matched aperture photometry was performed on the IRAC maps using \textsc{SExtractor} 
\citep{BertinArnouts_1996} in dual image mode, using the \chone channel as the detection band. We measure 
fluxes through $4\arcsec$ diameter circular apertures, which are corrected to estimates of total magnitude 
using aperture corrections (as measured by \citealt{Barmby_2008}) of $-0.35\pm0.04$, $-0.37\pm0.04$ mag in 
the 3.6\,$\micron$, 4.5\,$\micron$ channels respectively. The photometric uncertainties were scaled upwards by factors of 
2.8, 2.6 in the 3.6\,$\micron$, 4.5\,$\micron$ channels respectively, in order to account for noise correlation between pixels
introduced in the production of the mosaics which is not taken into account in the \textsc{SExtractor} error
estimates. These scaling factors were determined using the method outlined in \citet{Barmby_2008}. Finally, 
the uncertainties in the aperture corrections were added to the photometric errors in quadrature. 

The BCGs are extended objects for which the aperture corrected point source magnitudes are not 
good approximations to the total flux. We therefore use the \textsc{SExtractor} MAG\_AUTO magnitudes, again 
scaling up the photometric uncertainties using the method of \citet{Barmby_2008}. Note that we do not attempt
to deblend the BCGs beyond the level provided by \textsc{SExtractor}. While it is possible to use more sophisticated
techniques to improve deblending using higher resolution imaging at other wavelengths 
\citep[e.g., TFIT;][]{Laidler_2007}, we only possess relatively shallow ground-based optical data for the 
objects in our sample, which would require a large $k$-correction to IRAC wavelengths.

\begin{table}
\caption{Distance modulus (DM$_{\rm corr}$) and $k$-corrections applied to 
cluster photometry to correct each object to the median
redshift of the appropriate subsample. The correction applied is: corrected magnitude = observed apparent magnitude +$k$ + DM$_{\rm corr}$,
where $k$ is the $k$-correction in either the 3.6 or 4.5\,$\micron$ bands ($k_{3.6}$, $k_{4.5}$ respectively). See Section~\ref{s_LFMethod}
for details.}
\label{t_kCorr}
\begin{tabular}{|c|c|c|c|c|}
\hline
Cluster & $z$ & $k_{3.6}$ & $k_{4.5}$ & DM$_{\rm corr}$\\
\hline
\multicolumn{5}{l}{Low redshift subsample (median z = 0.42):}\\
ACT-CL J0509-5341 & 0.461 & +0.102 & +0.054 & -0.231\\
ACT-CL J0330-5227 & 0.442 & +0.053 & +0.028 & -0.121\\
ACT-CL J0438-5419 & 0.421 & +0.000 & +0.000 & +0.000\\
ACT-CL J0215-5212 & 0.480 & +0.151 & +0.080 & -0.337\\
ACT-CL J0304-4921 & 0.392 & -0.076 & -0.041 & +0.184\\
ACT-CL J0237-4939 & 0.334 & -0.202 & -0.125 & +0.590\\
ACT-CL J0235-5121 & 0.278 & -0.304 & -0.215 & +1.056\\
\\
\multicolumn{5}{l}{High redshift subsample (median z = 0.68):}\\
ACT-CL J0346-5438 & 0.530 & -0.329 & -0.182 & +0.673\\
ACT-CL J0102-4915 & 0.870 & +0.245 & +0.340 & -0.642\\
ACT-CL J0616-5227 & 0.684 & +0.000 & +0.000 & +0.000\\
ACT-CL J0528-5259 & 0.768 & +0.115 & +0.140 & -0.308\\
ACT-CL J0546-5345 & 1.066 & +0.436 & +0.690 & -1.188\\
ACT-CL J0232-5257 & 0.556 & -0.262 & -0.151 & +0.546\\
ACT-CL J0559-5249 & 0.609 & -0.140 & -0.097 & +0.306\\

\hline
\end{tabular}
\end{table}

\section{Infrared Luminosity Functions}
\label{s_LFs}
The luminosity function (LF) encodes key information about galaxy populations, and has
been used to characterise the variation of galaxy properties with environment \citep[e.g.,][]{Blanton_2003, 
DePropris_2003, Croton_2005, Popesso_2006, Loveday_2012} and redshift 
\citep[e.g.,][]{DePropris_1999, Lin_2006, Muzzin_2008, Mancone_2010, Capozzi_2012}. Since the IRAC
photometry probes the peak of the stellar light, the
IR LF serves as a good proxy for the stellar mass function. In this section, we present the first measurements
of the IR galaxy luminosity functions of SZ-selected clusters, and compare our results to those found from 
IR-selected cluster samples.

\begin{figure*}
\includegraphics[width=15cm]{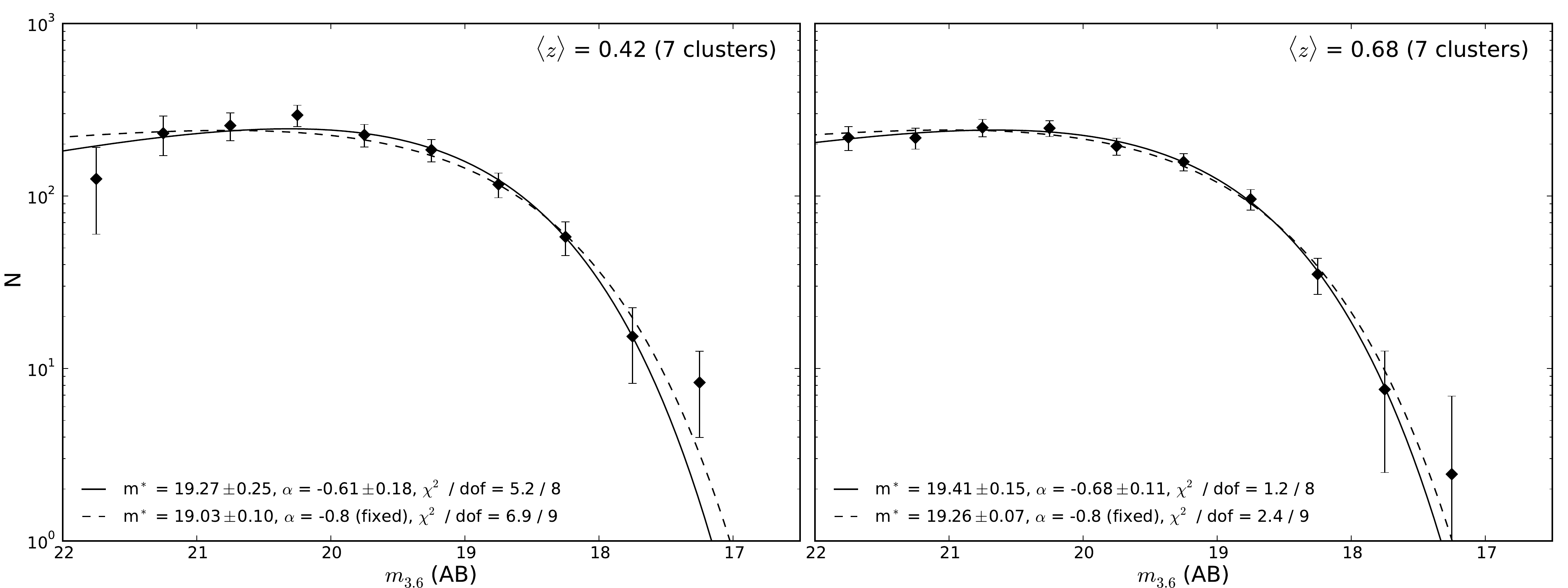}
\includegraphics[width=15cm]{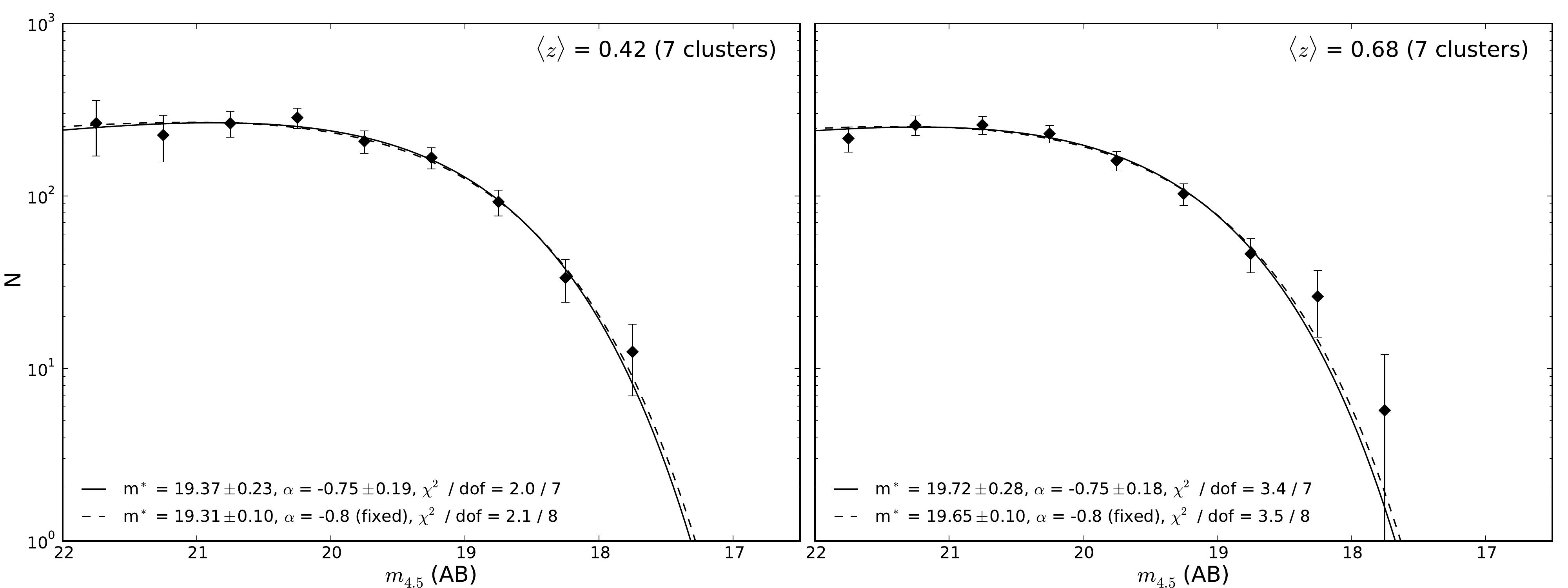}
\caption{Composite luminosity functions of ACT Sunyaev-Zel'dovich effect selected galaxy
clusters in two redshift bins, for each IRAC channel. The solid curve shows the best fitting Schechter function with $m^*$ and 
$\alpha$ as free parameters, while the dashed curve shows the best fit with $\alpha$ fixed at $-0.8$.}
\label{f_LFs}
\end{figure*}

\subsection{Method}
\label{s_LFMethod}
We divide the cluster sample by redshift into two equal sized subsamples ($0.2 < z < 0.5$ and 
$0.5 < z < 1.1$) and measure their composite luminosity functions within $R_{500}$.
Although we have a large sample of
spectroscopic redshifts for each cluster in addition to optical photometry, in most cases the optical data
do not provide coverage out to $R_{500}$. Therefore we use statistical background subtraction to ensure consistency
in our analysis across all the objects in the sample. This lack of optical data for some regions of most of the cluster 
fields, coupled to the low resolution of the IRAC images, makes star-galaxy separation difficult. Following 
\citet{Lin_2012}, we remove the brightest stars by cross-matching the IRAC catalogues with the 2MASS point 
source catalogue (using a 2$\arcsec$ matching radius); this should reduce the stellar contamination to around
the 7 per cent level \citep{Lin_2012}. We also remove all objects brighter than the BCG in each field, which
further reduces stellar contamination. Remaining stars are removed statistically during subtraction of the 
background sample. Since our IRAC cluster images do not extend to a large radius beyond $R_{500}$, we use the
IRAC source catalogue of the Extended Groth Strip \citep[EGS;][]{Barmby_2008} as the background field sample. 
We treat the EGS catalogue in the same manner as our IRAC catalogues of the cluster fields throughout (including,
for example, removing most stellar contamination by cross matching with 2MASS). We mask
out areas around very bright stars in some of our cluster fields, and take the reduction in area into account
when subtracting the background contribution.

We compute the composite luminosity functions for each subsample using the method of \citet{Colless_1989}.
We use the aperture corrected magnitudes (see Section~\ref{s_irac}) in the observed frame, and exclude the
BCGs since they are not drawn from the same population as ordinary cluster galaxies 
\citep[e.g.,][]{TremaineRichstone_1977, LohStrauss_2006, Lin_2010}. We apply a modest 
completeness correction (approximately 10 per cent for $21.5 < m_{3.6} < 22.0$), estimated from inserting synthetic 
point sources into our images (see Section~\ref{s_irac}). For the background sample from EGS, which reaches
similar depth to our observations, we use Table~5 of \citet{Barmby_2008} to model the completeness as a 
function of magnitude. For each cluster in each subsample, we $k$-correct its magnitudes (assuming a 
$\tau = 0.1$~Gyr single-burst \citet[][BC03 hereafter]{BruzualCharlot_2003} model, formed at $z_f = 3$ with
solar metallicity) to the median redshift of the subsample, and take into account the distance modulus between
each cluster and that median redshift \citep[see, for example, ][]{Muzzin_2008, Mancone_2010, Lin_2012}. 
The $k$ and distance modulus corrections adopted for each cluster are listed in Table~\ref{t_kCorr}.
We divide the data into bins of width 0.5 mag, and then perform the same
operations on the background galaxy sample, before subtracting the area-scaled background contribution in 
each bin.


We fit \citet{Schechter_1976} functions to the composite LFs using $\chi^2$ minimisation. We fit for both the 
characteristic magnitude ($m^*$) and the faint-end slope ($\alpha$), although we also perform fits with 
$\alpha$ fixed to other values common in the literature (e.g., $\alpha = -0.8$),
in order to simplify the comparison with other works (since $m^*$ and $\alpha$ are degenerate). The 
normalisation of the Schechter function fit is fixed such that its integral is equal to the number of galaxies in the composite LF.

\subsection{Results}
\begin{figure}
\includegraphics[width=8.5cm]{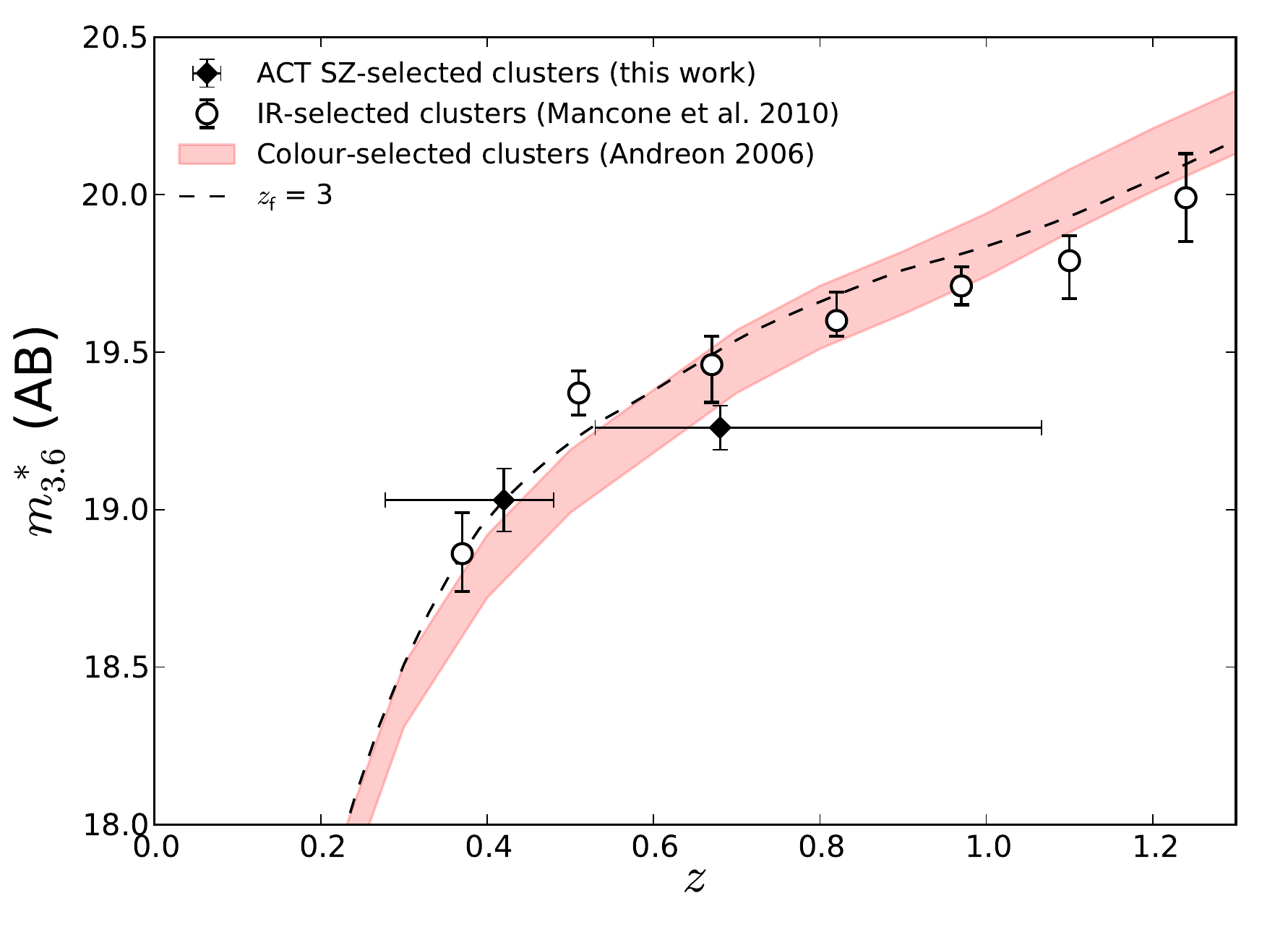}
\caption{Comparison of $m^*_{3.6}$ found from the SZ-selected cluster sample used in this work with 
results for an IR-selected cluster sample \citep{Mancone_2010}. In both cases, the faint end slope is fixed 
to $\alpha = -0.8$. There is reasonable agreement, which suggests that
the galaxy populations in each sample are not likely to be significantly different, although $m^*_{3.6}$ for
the $\langle z \rangle = 0.68$ subsample of SZ-selected clusters is approximately 0.3 mag brighter. The dashed
line shows the expected evolution of a solar metallicity \citet{BruzualCharlot_2003} $\tau=0.1$\,Gyr single 
burst stellar population model formed at $z_{\rm f} = 3$, normalised to the \citet{Mancone_2010} measurement
at $z = 0.37$. The horizontal error bars mark the redshift ranges covered by the ACT subsamples.}
\label{f_LFEvo}
\end{figure}

Fig.~\ref{f_LFs} shows the LFs in each IRAC channel for both the low and high redshift subsamples. The values
of $m^*$ that we derive in both channels for the $\langle z \rangle = 0.42$ subsample agree at better than
1$\sigma$ with measurements of $m^*$ in IR-selected cluster samples by \citet{Muzzin_2008} and 
\citet{Mancone_2010} at similar redshift (note that the faint end slope, $\alpha$, was fixed to -0.8 in these
works). We find a slightly brighter $m^*$ for the $\langle z \rangle = 0.68$ subsample ($m^*_{3.6} = 19.26 
\pm 0.07$ for $\alpha = -0.8$) in comparison to both \citet{Muzzin_2008} and \citet{Mancone_2010}, although
all values are consistent at better than the 2$\sigma$ level. If this difference is real, it may reflect a 
different timescale for the build up of the bright end of the luminosity function between the samples, with 
more massive galaxies being assembled at earlier times in more massive clusters (since we 
expect the average mass of the SZ-selected sample to be significantly larger than that of the IR-selected 
samples, which are drawn from a smaller survey area). Fig.~\ref{f_LFEvo} illustrates the evolution of 
$m^*_{3.6}$ in the \citet{Mancone_2010} sample in comparison to our results. We also see reasonable agreement
between $m^*$ measured for our $\langle z \rangle = 0.68$ subsample and the results of 
\citet{DePropris_2007} at $\langle z \rangle = 0.75$, when we set $\alpha = -0.25$, as was found in that work. 

The faint-end slopes that we find when fitting for both $m^*$ and $\alpha$ are consistent with that measured
by \citet{Lin_2004} in the $K$-band from a heterogeneous sample of X-ray selected clusters at $z = 0.1$, who 
found $\alpha = -0.84 \pm 0.02$, although we note that the uncertainties on our measurements are much larger.
We conclude that the infrared LFs of SZ-selected clusters do not differ significantly from clusters selected 
using other methods.

\section{Brightest Cluster Galaxies}
\label{s_BCGs}

\begin{figure}
\includegraphics[width=8.5cm]{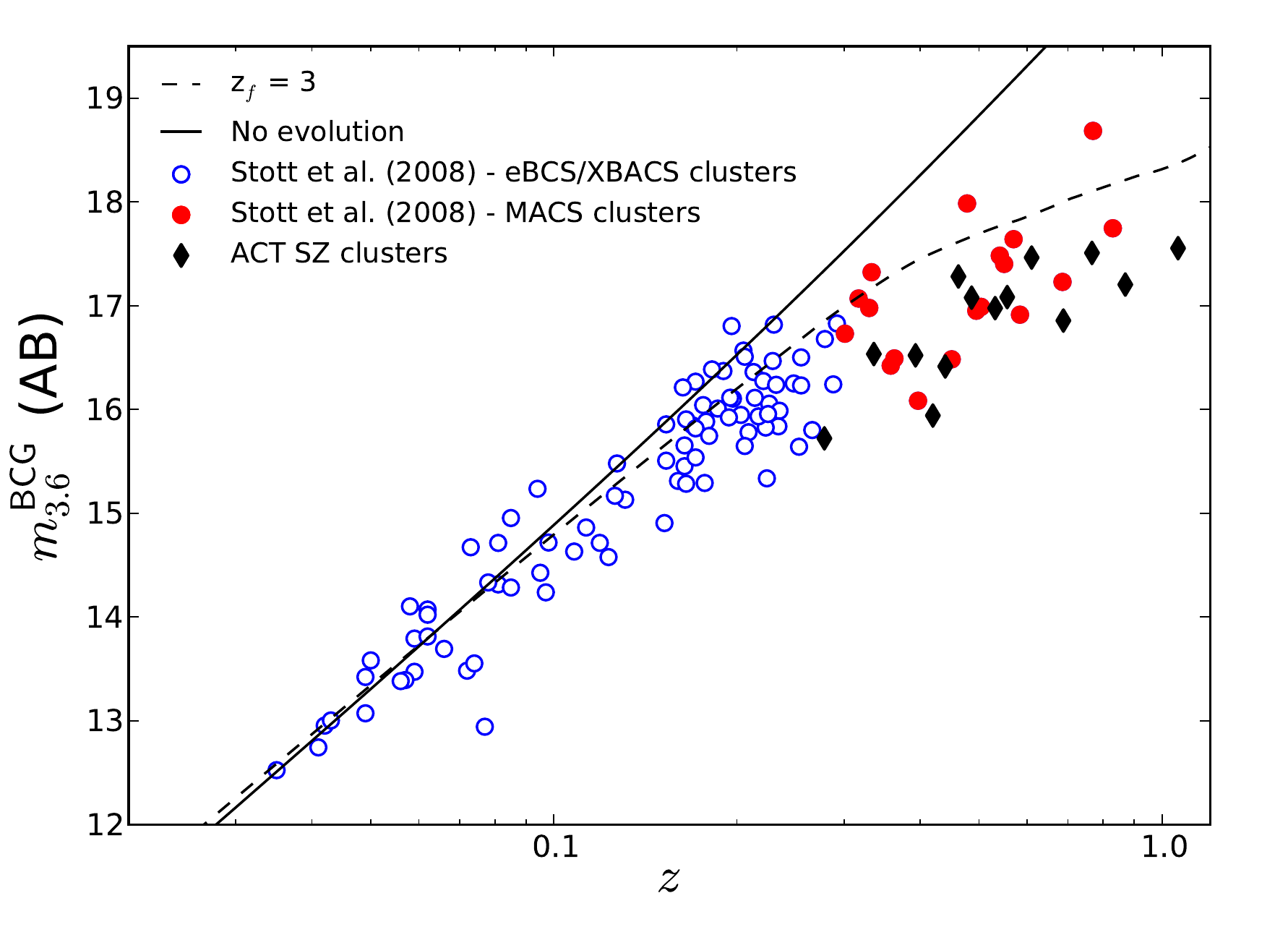}
\caption{Hubble diagram of BCGs in ACT clusters (black diamonds) in the 3.6\,$\micron$ IRAC channel. The $K$-band
observations ($k$-corrected to the IRAC 3.6\,$\micron$ channel assuming a $\tau = 0.1$\,Gyr
single burst, solar metallicity, \citet{BruzualCharlot_2003} model formed at $z_f = 3$) of two samples of X-ray selected clusters 
\citep{Ebeling_1996, Ebeling_2000, Ebeling_2001} by \citet{Stott_2008} are shown for comparison. The dashed line shows 
the expected evolution for the same \citet{BruzualCharlot_2003} model
used to perform the $k$-correction, while the solid line shows the no-evolution line; both of these are normalised to match the
data at $z < 0.1$.}
\label{f_BCGHubble}
\end{figure}

\begin{table*}
\caption{Properties of the BCGs; $m_{3.6}$ and $m_{4.5}$ are the total apparent magnitudes (observed frame) 
on the AB system in the IRAC 3.6 and 4.5\,$\micron$ channels respectively; $M_*$ is the stellar mass estimated
from $m_{3.6}$ alone, assuming a $\tau = 0.1$ Gyr burst \citet{BruzualCharlot_2003} model with 
\citet{Salpeter_1955} IMF and solar metallicity formed at $z_f = 3$. The error bars on the stellar mass 
estimates correspond to assumptions of $z_f = 2$ and $z_f = 5$ (systematic errors due to the choice of stellar
population model and/or IMF are neglected).}
\label{t_BCGs}
\begin{tabular}{|c|c|c|c|c|c|}
\hline
Cluster & RA (J2000) & Dec. (J2000) & $m_{3.6}$ & $m_{4.5}$ & $M_*$ ($10^{11}$\,M$_{\sun}$)\\
\hline
ACT-CL J0102-4915 & $01^{\rmn h} 02^{\rmn m} 57\fs772$ & $-49\degr 16\arcmin 19\farcs14$ & $17.204 \pm 0.009$ & $17.628 \pm 0.010$ & $16.2^{+2.8}_{-3.4}$\\
ACT-CL J0215-5212 & $02^{\rmn h} 15^{\rmn m} 12\fs229$ & $-52\degr 12\arcmin 25\farcs09$ & $17.078 \pm 0.006$ & $17.349 \pm 0.006$ & $10.9^{+0.6}_{-1.2}$\\
ACT-CL J0232-5257 & $02^{\rmn h} 32^{\rmn m} 42\fs704$ & $-52\degr 57\arcmin 22\farcs62$ & $17.083 \pm 0.006$ & $17.453 \pm 0.007$ & $12.1^{+1.2}_{-1.3}$\\
ACT-CL J0235-5121 & $02^{\rmn h} 35^{\rmn m} 45\fs242$ & $-51\degr 21\arcmin 04\farcs83$ & $15.722 \pm 0.002$ & $15.906 \pm 0.002$ & $17.6^{+1.1}_{-1.2}$\\
ACT-CL J0237-4939 & $02^{\rmn h} 37^{\rmn m} 01\fs661$ & $-49\degr 38\arcmin 09\farcs66$ & $16.535 \pm 0.005$ & $16.734 \pm 0.005$ & $11.3^{+0.7}_{-0.8}$\\
ACT-CL J0304-4921 & $03^{\rmn h} 04^{\rmn m} 16\fs132$ & $-49\degr 21\arcmin 25\farcs97$ & $16.522 \pm 0.004$ & $16.724 \pm 0.004$ & $14.4^{+0.8}_{-1.4}$\\
ACT-CL J0330-5227 & $03^{\rmn h} 30^{\rmn m} 56\fs935$ & $-52\degr 28\arcmin 13\farcs18$ & $16.415 \pm 0.004$ & $16.639 \pm 0.004$ & $18.3^{+1.0}_{-2.0}$\\
ACT-CL J0346-5438 & $03^{\rmn h} 46^{\rmn m} 55\fs370$ & $-54\degr 38\arcmin 54\farcs66$ & $16.977 \pm 0.006$ & $17.261 \pm 0.006$ & $12.9^{+1.1}_{-1.3}$\\
ACT-CL J0438-5419 & $04^{\rmn h} 38^{\rmn m} 17\fs644$ & $-54\degr 19\arcmin 20\farcs42$ & $15.942 \pm 0.003$ & $16.159 \pm 0.003$ & $26.9^{+1.5}_{-2.7}$\\
ACT-CL J0509-5341 & $05^{\rmn h} 09^{\rmn m} 21\fs375$ & $-53\degr 42\arcmin 12\farcs79$ & $17.282 \pm 0.010$ & $17.554 \pm 0.010$ & \phantom{0}$8.6^{+0.5}_{-0.9}$\\
ACT-CL J0528-5259 & $05^{\rmn h} 28^{\rmn m} 05\fs332$ & $-52\degr 59\arcmin 53\farcs27$ & $17.509 \pm 0.008$ & $17.965 \pm 0.009$ & $11.1^{+1.3}_{-2.4}$\\
ACT-CL J0546-5345 & $05^{\rmn h} 46^{\rmn m} 37\fs729$ & $-53\degr 45\arcmin 31\farcs41$ & $17.554 \pm 0.007$ & $17.835 \pm 0.007$ & $13.4^{+3.3}_{-2.3}$\\
ACT-CL J0559-5249 & $05^{\rmn h} 59^{\rmn m} 43\fs230$ & $-52\degr 49\arcmin 27\farcs05$ & $17.464 \pm 0.008$ & $17.873 \pm 0.010$ & \phantom{0}$9.2^{+1.0}_{-1.3}$\\
ACT-CL J0616-5227 & $06^{\rmn h} 16^{\rmn m} 34\fs090$ & $-52\degr 27\arcmin 08\farcs94$ & $16.857 \pm 0.004$ & $17.374 \pm 0.004$ & $18.1^{+1.7}_{-3.0}$\\
\hline
\end{tabular}
\end{table*}

\subsection{The Hubble diagram}
BCGs are the brightest galaxies in the Universe in terms of their stellar emission, and early work recognised 
their potential as standard candles, using their Hubble diagram to estimate the deceleration parameter 
\citep[e.g.,][]{Sandage_1972}. More recently, the Hubble diagram of BCGs has been used to study their mass 
growth \citep[e.g.,][]{AragonSalamanca_1998, Burke_2000, Brough_2002}. Some semi-analytic models predict that BCGs 
should have acquired about 80 per cent of their stellar mass since $z \sim 1$ through accretion and merging 
\citep{DeLucia_2007}, although recent observations of BCGs in both high redshift optical and X-ray 
selected clusters show that most of the stellar mass in these objects was already assembled by $z \sim 1$ 
\citep[e.g.,][]{Whiley_2008, Brough_2008, Collins_2009, Stott_2010, Lidman_2012}.

Fig.~\ref{f_BCGHubble} presents the \chone Hubble diagram of BCGs in the ACT SZ-selected sample, in comparison
to $K$-band observations of X-ray selected clusters ($k$-corrected to 3.6\,$\micron$, assuming a $\tau = 0.1$\,Gyr single burst, solar 
metallicity, BC03 model formed at $z_f = 3$), taken from \citet[][who measure total BCG magnitudes, as in 
this work]{Stott_2008}. One of 
the X-ray samples to which we compare is the Massive Cluster Survey
\citep[MACS;][]{Ebeling_2007, Ebeling_2010}, which is extracted from the ROSAT All Sky Survey 
\citep[][]{Voges_1999}, and probes a similar mass and redshift range to the ACT sample. Although we have 
normalised the evolution tracks shown in this figure to match the low redshift BCG sample, it should
be noted that the massive ACT clusters will not evolve to have similar masses to the low redshift clusters.

To check if the ACT and MACS BCGs trace similar populations, we convert our estimates of 
3.6\,$\micron$ BCG total magnitude to stellar mass, assuming a solar metallicity 
$\tau = 0.1$\,Gyr single burst BC03 model, formed at $z_f = 3$, and do the same for the MACS BCGs. We compare
the resulting BCG stellar mass distributions using the two-sample Kolmogorov-Smirnov (KS) test. When comparing to the 
whole MACS sample, the KS test returns $D = 0.44$, with null hypothesis (that the samples are drawn from the 
same distribution) probability $p = 0.06$. Restricting both samples to include only the $0.4 < z < 1.0$ BCGs, 
the null hypothesis probability increases to $p = 0.4$. We conclude that the ACT and MACS BCGs are a similar population.

\subsection{BCG stellar mass scaling relations}
\label{s_BCGMasses}
We now examine the scaling of BCG stellar mass with the properties of the ACT clusters. We convert our 
estimates of 3.6\,$\micron$ BCG total magnitudes to stellar mass, assuming a solar metallicity 
$\tau = 0.1$\,Gyr single burst BC03 model, formed at $z_f = 3$. The uncertainty in the
stellar mass estimates is dominated by the choice of model; since formation redshifts in the range $2 < z_f < 5$
are reasonable for BCGs \citep[e.g.,][]{Whiley_2008, Stott_2008, Stott_2010}, we adopt the stellar mass 
estimates inferred from models at each end of this redshift range as fiducial error bars (the impact of
photometric uncertainties is negligible by comparison). We do not take into
account possible systematic uncertainties, which are considerable. The largest systematic that can affect 
stellar mass estimates is the IMF; if we adopted a \citet{Chabrier_2003} rather than \citet{Salpeter_1955} 
IMF, our stellar mass estimates would be 0.24 dex lower. Similarly, adopting stellar population models with
a larger contribution to the infrared flux by thermally pulsating AGB stars \citep*[e.g.,][]{Maraston_2005,
Conroy_2009} would result in smaller stellar masses. Table~\ref{t_BCGs} lists the BCG magnitudes and stellar 
mass estimates derived using our adopted BC03 model.

\begin{figure}
\includegraphics[width=8.5cm]{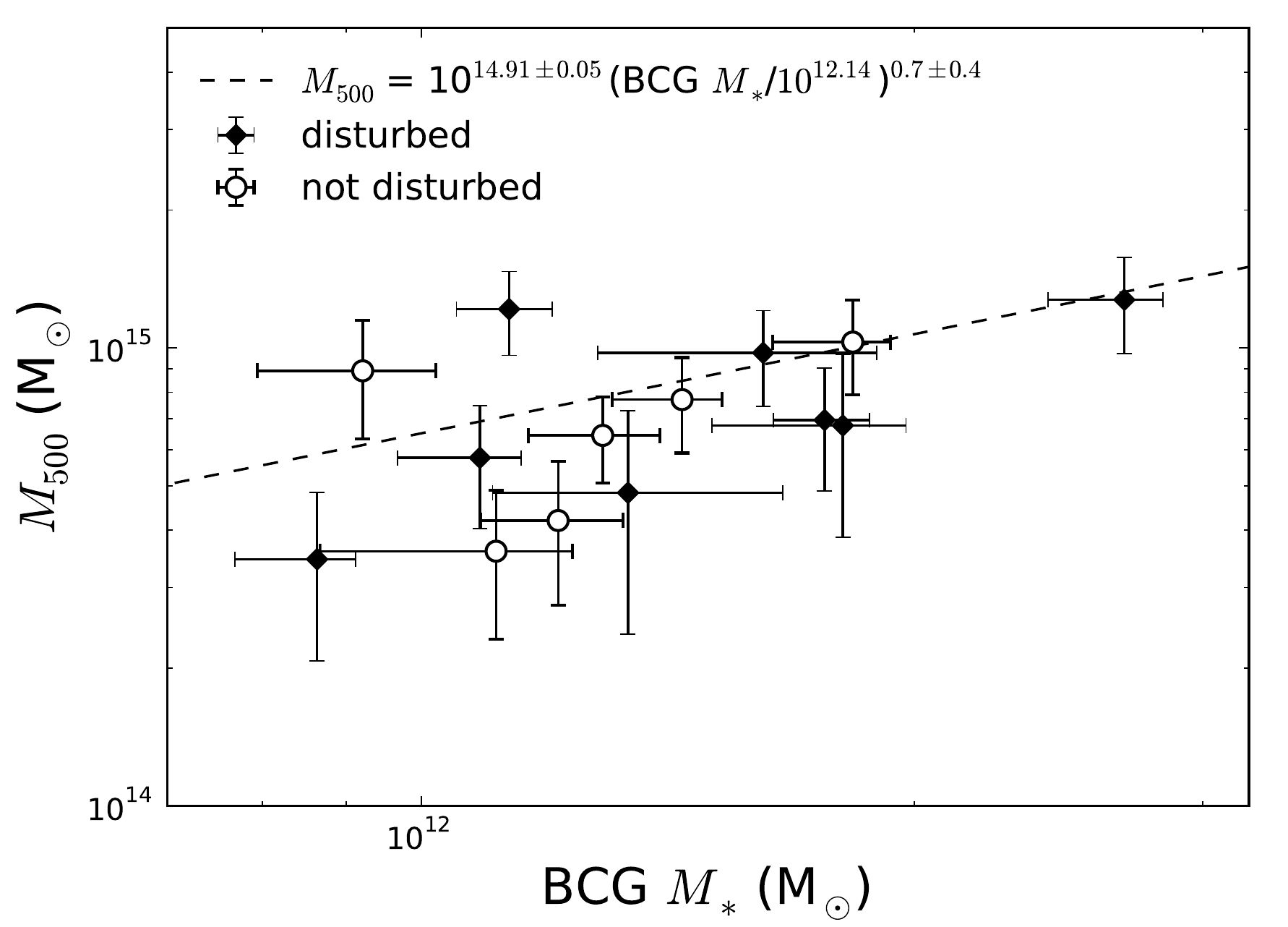}
\caption{Scaling relation between BCG stellar mass and cluster dynamical mass. The dashed line is a regression
fit to the data using the \citet{Kelly_2007} method.
There is no evidence from this small sample that disturbed clusters (black diamonds; defined according to the 
criteria described in \citealt{Sifon_2012}) follow a different trend from non-disturbed clusters (white circles).}
\label{f_BCGScaling_M500}
\end{figure}

Fig.~\ref{f_BCGScaling_M500} presents the relation between BCG stellar mass and cluster dynamical mass. We find
mild evidence for a correlation, with Spearman rank coefficient $\rho = 0.56$, and null hypothesis (i.e.
no correlation) probability $p = 0.04$. Using the Markov Chain Monte Carlo (MCMC) based method of 
\citet{Kelly_2007}, we find the slope of the relation is shallow, and poorly constrained 
($M_{500} \propto M_*^{0.7 \pm 0.4}$). The intrinsic scatter in the relation at fixed BCG $M_*$ is 
$\sigma_{\log M_{500}} = 0.14 \pm 0.06$.

Many other studies have previously found a correlation between 
BCG luminosity and cluster mass using X-ray and optically selected cluster samples 
\citep[e.g.,][]{LinMohr_2004, Popesso_2007, Brough_2008, Yang_2008, Mittal_2009}. 
Like \citet{Haarsma_2010}, the correlation we find is only marginally significant; presumably this due to the 
small sample size. For example, \citet{Whiley_2008} found a correlation between BCG $K$-band 
magnitude and cluster velocity dispersion (i.e. a comparable proxy to the dynamical mass estimates used in 
this work), significant at the 99.9 per cent level, for a sample of 81 optically selected clusters in the
$0.02 < z < 0.96$ redshift range. We note that the wide redshift range covered by our sample may also be a 
contributing factor, if the relation evolves with redshift.

\begin{figure}
\includegraphics[width=8.5cm]{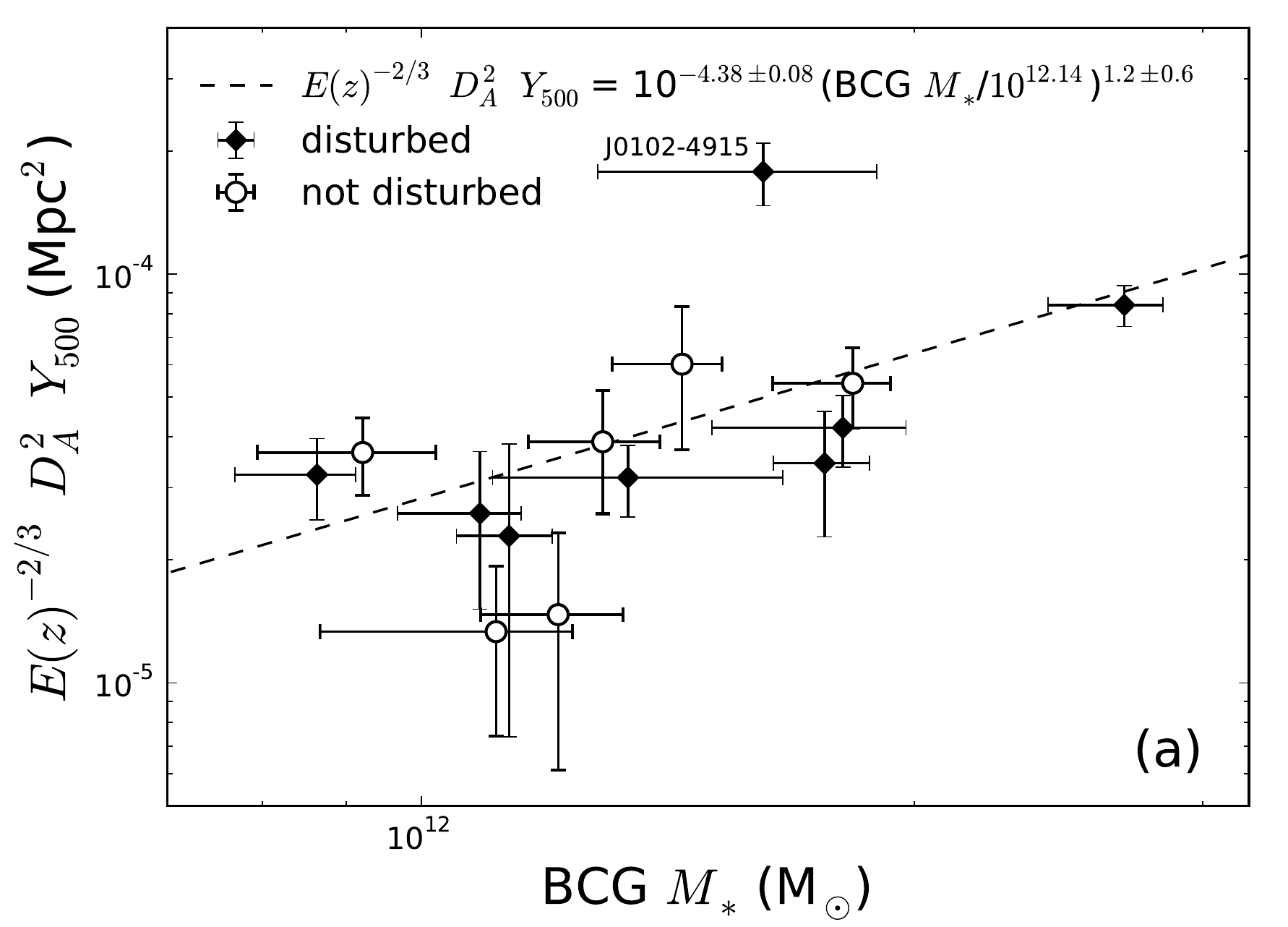}
\includegraphics[width=8.5cm]{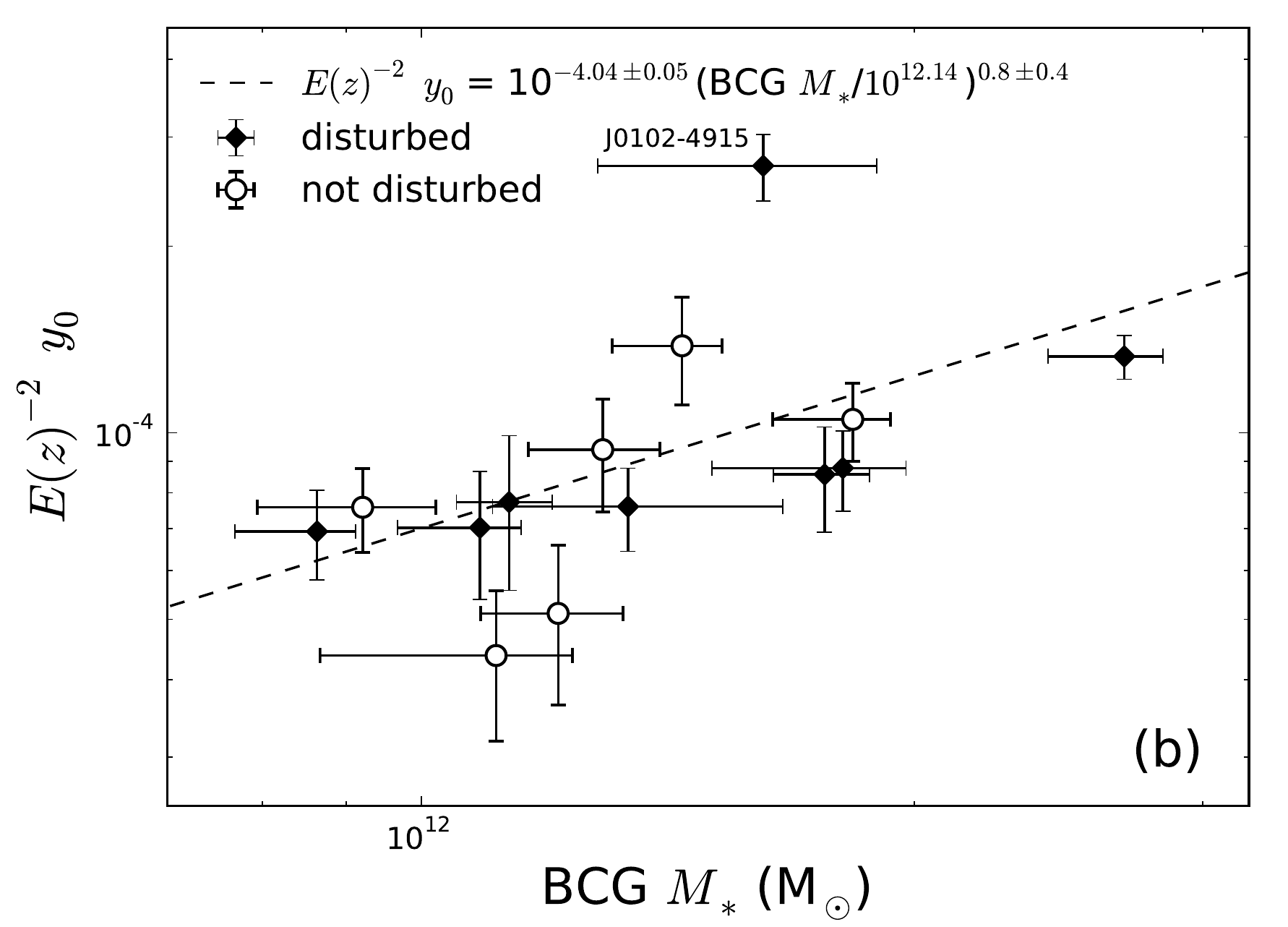}
\caption{Scaling relations between SZ-signal and BCG stellar mass: (a) scaling with $Y_{500}$; 
(b) scaling with $y_0$. Self-similar evolution of the SZ-signal with redshift is assumed.
In each panel the dashed line is a fit to the data using the \citet{Kelly_2007} regression method. 
We do not see different trends for disturbed (black diamonds; defined according to the criteria 
described in \citealt{Sifon_2012}) and non-disturbed clusters (white circles).}
\label{f_BCGScaling_SZ}
\end{figure}

Fig.~\ref{f_BCGScaling_SZ} shows the relations between BCG stellar mass and SZ-signal, $Y_{500}$ and
$y_0$. These are the first measurements of these relations. In these cases we find stronger evidence for
correlation with the BCG stellar mass. For the scaling with $Y_{500}$, we find Spearman rank coefficient 
$\rho = 0.66$, with null hypothesis probability $p = 0.01$. Applying the \citet{Kelly_2007} regression 
method, the slope of the relation is $E(z)^{-2/3}\,D_A^2\,Y_{500} \propto M_*^{1.2 \pm 0.6}$, with 
intrinsic scatter $\sigma_{\log{Y_{500}}} = 0.25 \pm 0.08$. For the BCG $M_*-y_0$ relation, we find 
Spearman $\rho = 0.74$, with $p = 0.002$. The slope of the relation is 
$E(z)^{-2}\,y_0 \propto M_*^{0.8 \pm 0.4}$, and the intrinsic scatter is 
$\sigma_{\log{y_{0}}} = 0.18 \pm 0.05$. We note that ``El Gordo'' (J0102-4915), the cluster with the largest $y_0$
and $Y_{500}$ value, is a clear outlier in these plots. However, removing this object has no significant effect on 
the fit results.

The extensive spectroscopic data obtained on each cluster allows their dynamical states to be classified 
using three different methods, as described in \citet{Sifon_2012}. A cluster is flagged as dynamically 
disturbed if it satisfies any two of (i) BCG peculiar velocity different from zero at more than the 2$\sigma$ level;
(ii) BCG projected offset from the SZ peak cluster position more than 0.2$R_{200}$; and (iii) greater than 5 
per cent significance level in the \citet{DresslerShectman_1988} test for substructure. In both 
Figs.~\ref{f_BCGScaling_M500}~and~\ref{f_BCGScaling_SZ} we show disturbed and non-disturbed clusters with 
different symbols, and see no evidence for different scaling of $M_{500}$, $Y_{500}$, or $y_0$ with BCG stellar
mass for clusters with different dynamical states, though of course the sample size is small.


\section{Cluster stellar mass scaling relations}
\label{s_totalStellarMass}

\begin{table}
\caption{Cluster total 3.6\,$\micron$ magnitudes (observed frame) and total stellar mass 
($M^{\rm star}_{500}$) estimated within $R_{500}$. The stellar mass conversion is done as described in 
Section~\ref{s_BCGMasses} for the BCGs, and the same caveats noted in the caption to Table~\ref{t_BCGs} apply.
Note that any contribution to $M^{\rm star}_{500}$ from the ICL is not accounted for.}
\label{t_stellarMasses}
\begin{tabular}{|c|c|c|c|}
\hline
Cluster & $z$ & $m_{3.6}$ & $M^{\rm star}_{500}$ ($10^{12}$\,M$_{\sun}$)\\
\hline
ACT-CL J0102-4915 & 0.870 & $14.22 \pm 0.09$ & $18.8^{+3.5}_{-4.1}$\\
ACT-CL J0215-5212 & 0.480 & $14.54 \pm 0.09$ & \phantom{0}$8.5^{+0.8}_{-1.1}$\\
ACT-CL J0232-5257 & 0.556 & $14.34 \pm 0.07$ & $11.4^{+1.5}_{-1.5}$\\
ACT-CL J0235-5121 & 0.278 & $12.85 \pm 0.03$ & $18.6^{+1.6}_{-1.6}$\\
ACT-CL J0237-4939 & 0.334 & $14.26 \pm 0.10$ & \phantom{0}$7.0^{+1.0}_{-1.0}$\\
ACT-CL J0304-4921 & 0.392 & $14.02 \pm 0.07$ & $10.9^{+1.0}_{-1.3}$\\
ACT-CL J0330-5227 & 0.442 & $13.44 \pm 0.07$ & $21.2^{+1.8}_{-2.6}$\\
ACT-CL J0346-5438 & 0.530 & $14.07 \pm 0.06$ & $14.0^{+1.6}_{-1.7}$\\
ACT-CL J0438-5419 & 0.421 & $13.30 \pm 0.06$ & $23.1^{+1.8}_{-2.7}$\\
ACT-CL J0509-5341 & 0.461 & $14.48 \pm 0.11$ & \phantom{0}$8.6^{+1.0}_{-1.2}$\\
ACT-CL J0528-5259 & 0.768 & $14.59 \pm 0.11$ & $12.2^{+1.8}_{-2.9}$\\
ACT-CL J0546-5345 & 1.066 & $14.50 \pm 0.14$ & $16.6^{+4.2}_{-3.1}$\\
ACT-CL J0559-5249 & 0.609 & $13.74 \pm 0.06$ & $21.1^{+2.7}_{-3.2}$\\
ACT-CL J0616-5227 & 0.684 & $13.94 \pm 0.08$ & $19.9^{+2.3}_{-3.6}$\\
\hline
\end{tabular}
\end{table}

In this Section, we examine the scaling relations between total stellar mass within $R_{500}$ 
($M_{500}^{\rm star}$), cluster dynamical mass, and SZ signal.

\subsection{Method}
We estimate the contribution to the total light within $R_{500}$ from 
cluster galaxies other than the BCG via the procedure used to measure the composite LFs (Section~\ref{s_LFs}). 
We sum the flux from the background-subtracted number counts for each cluster down to a limit of $m^*+2$ (where 
$m^*$ is the value obtained for the fit to the appropriate composite LF). We add a contribution for galaxies fainter than 
our detection limit by extrapolating the LF to $m^*+5$, assuming that $\alpha = -0.8$ 
(this correction adds less than 10 per cent to the final stellar masses). We then convert the total flux (adding 
in the contribution from the BCG, see Section~\ref{s_BCGs}) into an observed frame \chone magnitude, taking into 
account the difference in distance modulus and $k$-corrections that were applied in estimating the composite LF
(recall that we previously $k$-corrected the photometry for each cluster in a subsample to the median redshift of that 
subsample; see Table~\ref{t_kCorr}). The stellar mass is then estimated assuming the same $z_f = 3$ BC03 model applied in 
Sections~\ref{s_LFs} and 
\ref{s_BCGs}. Finally, to obtain spherically deprojected values, we multiply the total stellar mass estimates by 
0.73. This deprojection factor is obtained assuming an NFW profile \citep{NFW_1997} with $c = 2.8$ 
(the average obtained from applying the \citet{Duffy_2008} $c - M_{200}$ relation to the dynamical mass 
estimates), and integrating to 3$R_{500}$ along the line of sight. Note that our measurement does not include any
contribution from the intracluster light (ICL). Table~\ref{t_stellarMasses} lists the $M_{500}^{\rm star}$ value
for each cluster.

\begin{figure}
\includegraphics[width=8.5cm]{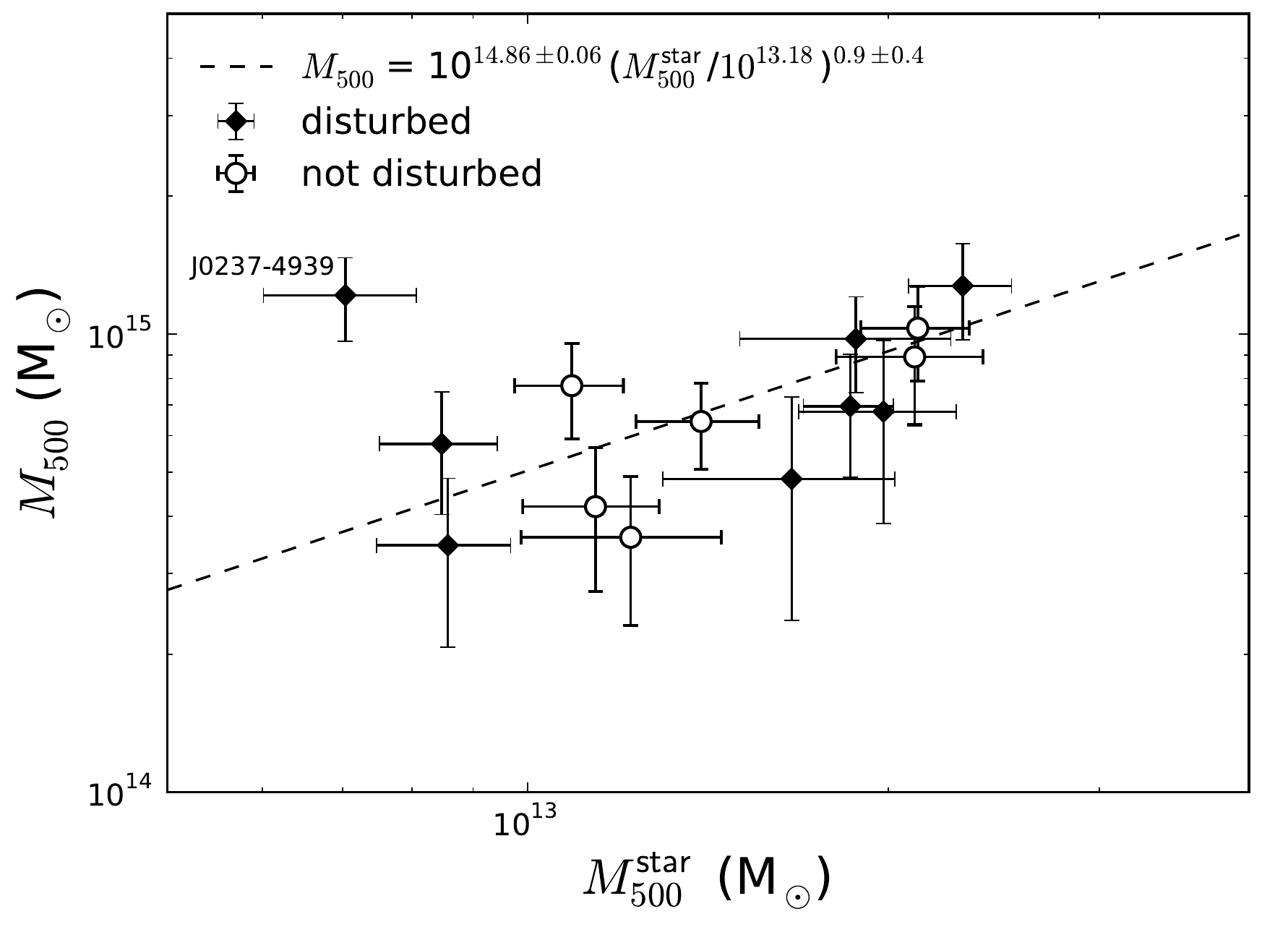}
\caption{Scaling relation between total stellar mass ($M^{\rm star}_{500}$) and dynamical cluster mass
($M_{500}$). The dashed line indicates the fit obtained with the \citet{Kelly_2007} regression method, after
excluding the outlier J0237-4939, which has a very low stellar mass compared to its dynamical mass. 
\textit{Chandra} X-ray observations of this cluster indicate its dynamical mass may be overestimated.}
\label{f_stellarMassM500}
\end{figure}

\begin{figure}
\includegraphics[width=8.5cm]{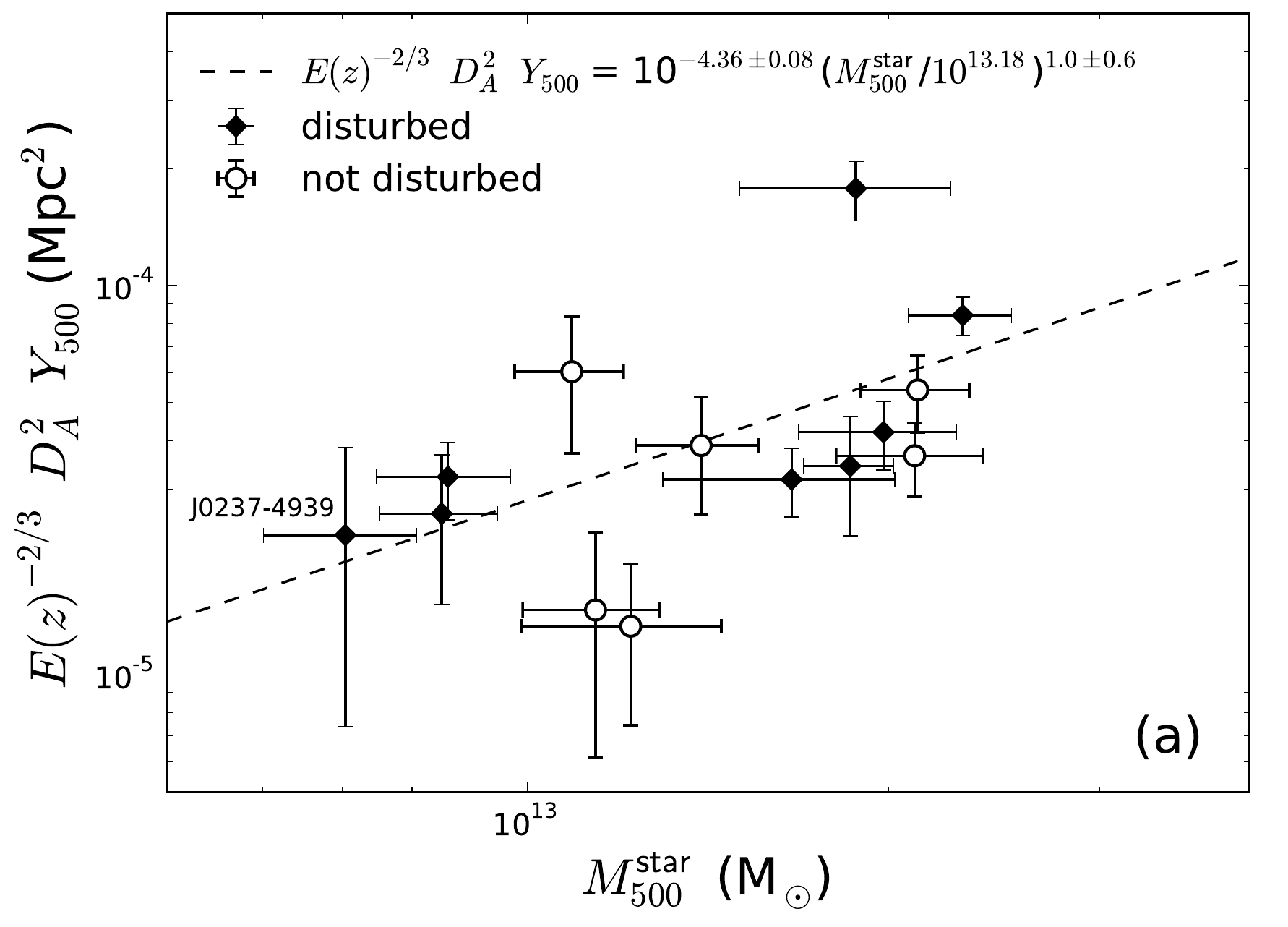}
\includegraphics[width=8.5cm]{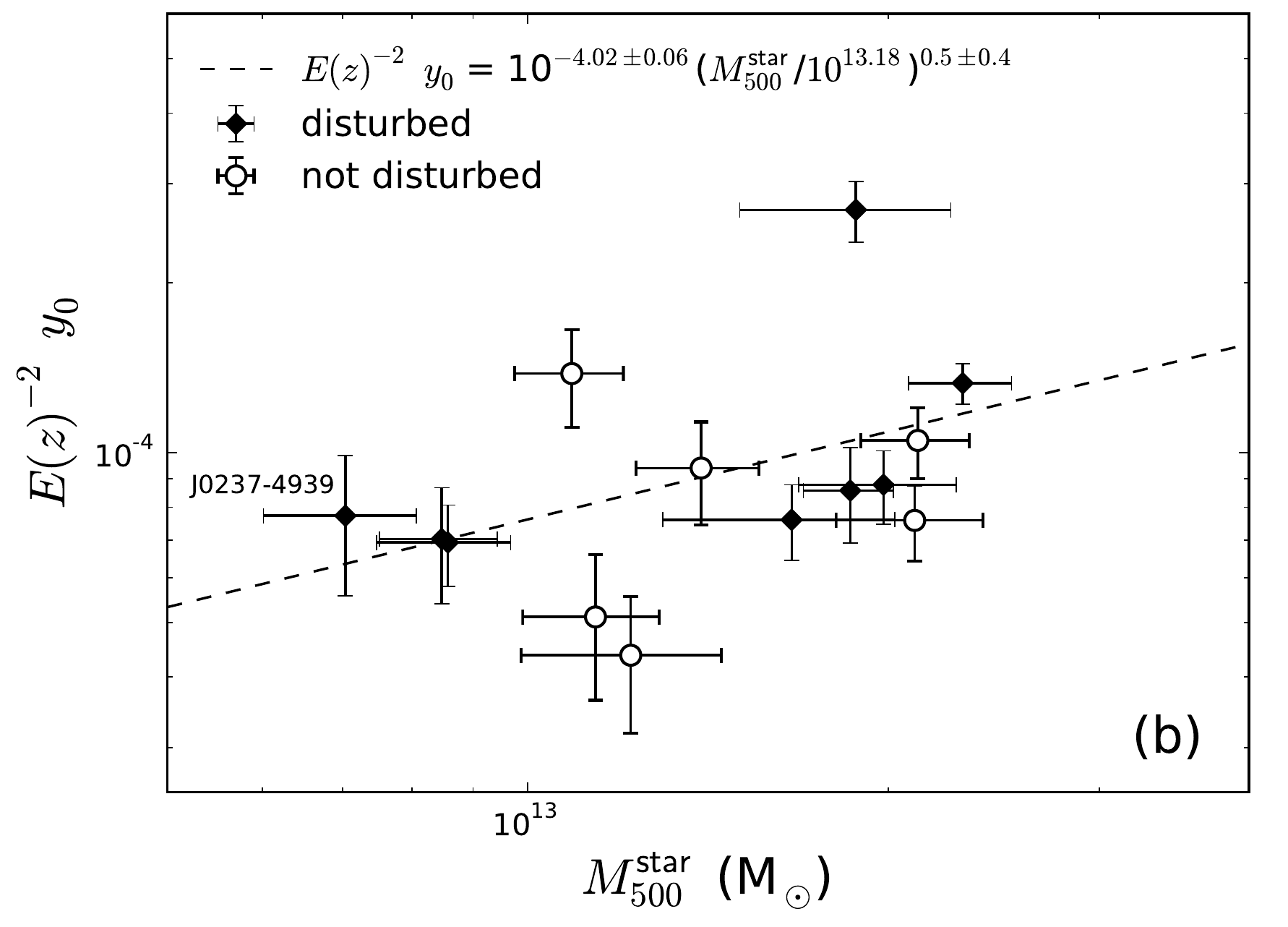}
\caption{Scaling relations between SZ-signal and cluster stellar mass: (a) scaling with $Y_{500}$; 
(b) scaling with $y_0$. Self-similar evolution is assumed in both cases. The dashed line indicates the
fit obtained using the \citet{Kelly_2007} regression method. J0237-4939, which has a very low stellar
mass compared to its dynamical mass (see Section~\ref{s_stellarMassResults}), is not an outlier in 
either relation.}
\label{f_stellarMassY200}
\end{figure}

\subsection{Results}
\label{s_stellarMassResults}
Fig.~\ref{f_stellarMassM500} shows the scaling of $M^{\rm star}_{500}$ with dynamical mass. While there is some
evidence of a correlation in this plot, we find that J0237-4939 is a clear outlier, as it has the smallest 
$M_{500}^{\rm star}$ in the sample but is the second ranked cluster in the sample in terms of its dynamical mass.
A preliminary analysis of the \textit{Chandra} X-ray data for this system indicates the dynamical mass is likely 
overestimated - the X-ray temperature is $T = 5.3^{+0.8}_{-0.7}$\,keV, which implies mass 
$M_{500} = (4.0 \pm 1.0) \times 10^{14}$\,M$_{\sun}$ (assuming the \citealt{Vikhlinin_2009} $M-T$ relation). This is
roughly one third of the dynamical mass for this system as listed in Table~\ref{t_clusterMasses}. A full 
comparison of dynamical versus X-ray derived mass estimates for the ACT sample will be presented in a future
paper (Hughes et al., in prep.).

If this cluster is excluded, we obtain Spearman rank coefficient $\rho = 0.76$ with null hypothesis 
probability $p = 0.002$. Applying the \citet{Kelly_2007} regression method with J0237 excluded, we find the 
relation $M_{500} \propto M_{500}^{\rm star}\,^{0.9 \pm 0.4}$, with intrinsic scatter 
$\sigma_{\log{M_{500}}} = 0.10 \pm 0.06$.
If J0237 is included in the sample, the correlation is not significant ($\rho = 0.47$, $p = 0.09$) and a much
shallower slope for the relation is inferred ($M_{500} \propto M_{500}^{\rm star}\,^{0.3 \pm 0.4}$).

We present the relations between $M^{\rm star}_{500}$ and SZ-signal ($y_0$, $Y_{500}$) in 
Fig.~\ref{f_stellarMassY200}.  In both cases we see a correlation, although the Spearman rank test indicates that
the correlation with $Y_{500}$ is more significant ($\rho = 0.63$, $p = 0.02$) compared to the correlation with
$y_0$ ($\rho = 0.46$, $p = 0.10$). In both cases, as with the scaling with the BCG stellar mass, the slopes of the
relations are poorly constrained ($E(z)^{-2/3}\,D_A^2\,Y_{500} \propto M_{500}^{\rm star}\,^{1.0 \pm 0.6}$; 
$E(z)^{-2}\,y_0 \propto M_{500}^{\rm star}\,^{0.5 \pm 0.4}$) with large intrinsic scatter 
($\sigma_{\log{Y_{500}}} = 0.26 \pm 0.09$; $\sigma_{\log{y_{0}}} = 0.20 \pm 0.06$). 

Despite the fact that J0237 is an outlier in the scaling of $M^{\rm star}_{500}$ with dynamical mass, it is not an outlier in either of the
SZ-signal relations. If we exclude it from the sample, we find less than 1$\sigma$ shifts in all of the fit
parameters and statistics quoted above for both the $y_0$--$M^{\rm star}_{500}$ and $Y_{500}$--$M^{\rm star}_{500}$ 
scaling relations.  

\section{Discussion}
\label{s_discuss}

We now discuss the variation of the baryon fraction in stars ($f^{\rm star}_{500} = M^{\rm star}_{500}/M_{500}$) 
with cluster mass for the ACT SZ-selected sample. This relationship may give some insight into the role of 
feedback in regulating star formation efficiency in clusters \citep[e.g.,][]{Bode_2009}. 

\begin{figure*}
\includegraphics[width=12cm]{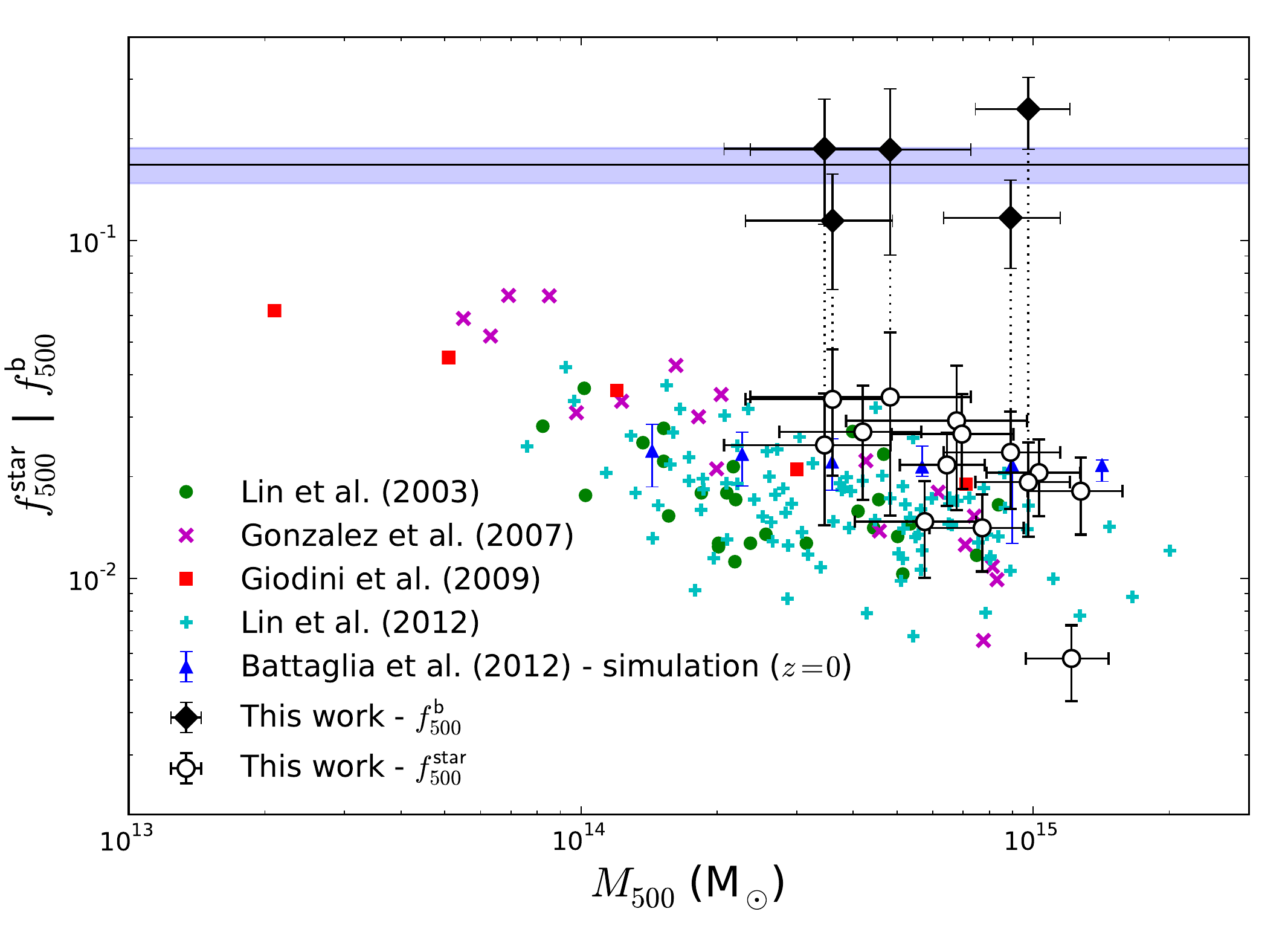}
\caption{Fraction of the total cluster mass in stars ($f^{\rm star}_{500}$) or total baryon fraction 
($f_{500}^{\rm b}$, i.e., $M^{\rm star}_{500}$ plus gas mass, where available) within $R_{500}$, as a function of
cluster mass. The solid line marks the cosmic baryon fraction 
\citep[][the shaded area indicates the uncertainty]{Komatsu_2011}. The legend indicates 
measurements of $f^{\rm star}_{500}$ from the literature (error bars are omitted for clarity); note that the 
results of \citet{Lin_2012} are scaled up from a \citet{Kroupa_2001}
IMF by adding 0.13 dex to the total stellar masses. Only the work of \citet{Gonzalez_2007} includes the 
contribution from the ICL. The five black diamonds indicate
$f_{500}^{\rm b}$ estimates for J0102-4915 \citep[gas mass measurement taken from][]{Menanteau_2012}, 
J0509-5341, J0528-5259, J0546-5345, and J0559-5249 \citep[gas mass measurements taken from][]{Andersson_2011}.}
\label{f_fBaryons}
\end{figure*}

Fig.~\ref{f_fBaryons} shows our results in comparison to a number of works in the literature. For the ACT sample,
we find $f^{\rm star}_{500}$ values that span the range 0.006--0.034, with median 0.022. The cluster at the low
end of this range is J0237-4939, which has an unusually low stellar mass given its dynamical mass (although as
noted above, analysis of the X-ray data for this system indicates the dynamical mass is overestimated). It is 
flagged as a disturbed cluster according to the criteria used by \citet[][note that this is also indicated by
the morphology of the cluster in \textit{Chandra} X-ray imaging]{Sifon_2012}, although the other 
disturbed clusters in the sample do not exhibit such low $f^{\rm star}_{500}$ values. In Fig.~\ref{f_fBaryons} we see the
decreasing trend of $f^{\rm star}_{500}$ with increasing cluster mass seen in other works, though
the mass range covered by the ACT sample is relatively narrow. The $f^{\rm star}_{500}$ values we find are 
typically slightly larger at a given mass in comparison to the results of other studies shown in the plot. While
this could be a real difference, there are several possible sources of systematic error, or differences in the
analyses or assumptions applied in each study, which could account for this.

The single largest source of systematic error is the choice of IMF. Here, we have adopted the 
\citet{Salpeter_1955} IMF (as did the work of \citealt{Giodini_2009}), and in Fig.~\ref{f_fBaryons} we have
rescaled the results of \citet{Lin_2012} from the \citet{Kroupa_2001} IMF assumed in that work to 
\citeauthor{Salpeter_1955} by adding 0.13~dex to
their $M_{500}^{\rm star}$ values. If we had instead adopted a \citet{Chabrier_2003} IMF instead of 
\citeauthor{Salpeter_1955}, our stellar mass measurements would be 0.24~dex lower. \citet{Gonzalez_2007} took a 
different approach, converting light to stellar mass using an empirically determined sub-Salpeter $M/L$ ratio 
taken from observations of local ellipticals and S0s
by \citet{Cappellari_2006}. We note that the possible variation of the IMF according to galaxy type is currently
under debate; some recent studies \citep[][]{vanDokkumConroy_2010, Cappellari_2012} suggest that early type 
galaxies have a more bottom-heavy IMF relative to late type galaxies. Accounting for such an effect is beyond the
scope of this paper.

In a similar way, the assumption made in converting light to mass is also a potential source of systematic 
error. In the absence of multi-band photometry covering the entire sample out to $R_{500}$, we have simply assumed
a constant mass-to-light ratio for each cluster, assuming all member galaxies to be represented by a single-burst 
BC03 model with $z_f = 3$. This is motivated by the fact that early type galaxies form the dominant population in
massive clusters, even beyond $z > 1$, with the scatter about the red-sequence in such clusters being consistent
with the bulk of star formation having taken place at $z \approx 3$ \citep[e.g.,][]{Mei_2009, Hilton_2009, 
Strazzullo_2010}. This assumption is also made in \citet{Lin_2012}, and there is reasonable agreement between
the two studies in the mass range of overlap. However, some other studies take into account the variation in 
mass-to-light ratio with galaxy type. \citet{Giodini_2009} apply the $K_s$-to-stellar mass relations of
\citet{Arnouts_2007}, which are derived from SED fitting to multiband photometry, according to galaxy colour. 
\citet{Leauthaud_2012} used SED fitting to COSMOS photometry to obtain stellar mass estimates for each individual
galaxy, and estimated the stellar fraction using a statistical halo occupation distribution model approach, 
finding much lower values (by a factor of 2-5)
than for other studies, including ours (although much of this difference can be attributed to the adoption
of the \citet{Chabrier_2003} IMF). We note that such a galaxy-type dependent estimation of stellar mass will have
a much larger impact at the group scale, the focus of the work by \citet{Giodini_2009} and \citet{Leauthaud_2012},
in comparison to the massive clusters we consider here.

As noted earlier, we do not account for the presence of ICL when computing $M_{500}^{\rm star}$. Of the studies
shown in Fig.~\ref{f_fBaryons}, only the work of \citet{Gonzalez_2007} measured and included the ICL component, 
which was found to be approximately 30 per cent of the total light across a sample of $z \approx 0.1$ clusters. 
Recent studies suggest that the ICL makes up a smaller fraction of the total cluster light at high redshift
\citep[$\approx 4$ per cent at $z = 0.8$;][]{Burke_2012}.

Despite the limitations of the analysis in this work, the stellar fractions we find are nevertheless in
reasonable agreement with the results of numerical simulations by \citet{Battaglia_2012}. These cosmological 
simulations include sub-grid models for radiative cooling, star formation, and AGN feedback 
\citep{Battaglia_2010}. As can be seen in Fig.~\ref{f_fBaryons}, our observations (in common with the other
observational results plotted) show a larger scatter than the simulations, which suggests that the 
sub-grid models in the simulations need to be improved to capture the larger observed variations in 
$f_{500}^{\rm star}\,(M_{500})$.

The stellar component of clusters makes only a small contribution to the total
baryonic mass. Measurements of gas mass within $R_{500}$ ($M_{500}^{\rm gas}$) are available from the literature
for five of the objects in the ACT sample \citep{Andersson_2011, Menanteau_2012}. In Fig.~\ref{f_fBaryons} we plot
the total baryon fractions ($f_{500}^{\rm b} = [M_{500}^{\rm star} + M_{500}^{\rm gas}]/M_{500}$) for these
clusters, in comparison to the cosmic mean as measured by WMAP \citep{Komatsu_2011}. We see no evidence for a 
shortfall of baryons in these objects given current uncertainties, although this is clearly dependent upon the 
validity of the assumptions made in estimating $f_{500}^{\rm star}$.

\section{Conclusions}
\label{s_conclusions}

We have performed the first study of the stellar mass component of galaxy clusters selected via the 
Sunyaev-Zel'dovich effect, using \textit{Spitzer Space Telescope} infrared observations of a sample 
detected by the Atacama Cosmology Telescope. We found:

\begin{enumerate}

\item The \chone and \chtwo luminosity functions are similar to those measured for IR-selected cluster samples.
We measure both a characteristic magnitude ($m^*$) and faint-end slope ($\alpha$) for the low 
redshift ($0.2 < z < 0.5$) ACT sample similar to those found by \citet{Mancone_2010}, while for the higher redshift
sample ($0.5 < z < 1.1$), the ACT clusters have slightly brighter $m^*$ (at the $2\sigma$ level).
\vskip 3pt

\item The relation between BCG stellar mass and cluster dynamical mass for the ACT sample is shallow
($M_{500} \propto M_*^{0.7 \pm 0.4}$). The relation is not very well constrained, 
most likely due to the small sample size. There is strong evidence for the correlation of BCG stellar mass
with SZ-observables, although again the 
constraints we obtained on the slopes of these relations are poor. For the scaling with $Y_{500}$, we found
$E(z)^{-2/3}\,D_A^2\,Y_{500} \propto M_*^{1.2 \pm 0.6}$, with intrinsic scatter 
$\sigma_{\log{Y_{500}}} = 0.25 \pm 0.08$.
\vskip 3pt

\item Excluding the cluster J0237-4939, which has anomalously low total stellar mass ($M_{500}^{\rm star}$) 
compared to its dynamical mass, we found $M_{500} \propto M_{500}^{\rm star}\,^{0.9 \pm 0.4}$, with 
intrinsic scatter $\sigma_{\log{M_{500}}} = 0.10 \pm 0.06$. We also made the first measurement of the scaling
of $M_{500}^{\rm star}$ with SZ-signal, finding 
$E(z)^{-2/3}\,D_A^2\,Y_{500} \propto M_{500}^{\rm star}\,^{1.0 \pm 0.6}$.
\vskip 3pt

\item The stellar fractions that we measured cover the range 0.006--0.034, with median 0.022. These are larger 
than found in some other studies of clusters of similar mass, but in reasonable agreement with the results of a similar
analysis of X-ray selected clusters using WISE data \citep{Lin_2012}. For the five clusters with additional gas
mass measurements available in the literature, we see no evidence for a shortfall of baryons in clusters relative
to the cosmic mean value.
\vskip 3pt
\end{enumerate}

In the future, we intend to extend this study to include ACT clusters on the celestial equator 
\citep[Hasselfield~et~al.~2013;][]{MenanteauEq_2012}, where the overlap with 
deep SDSS Stripe 82 optical photometry \citep{Annis_2012} will allow us to perform a more detailed analysis, 
without the need for some of the assumptions used in this work.

\section*{Acknowledgments}
This work is based in part on observations made with the \textit{Spitzer Space Telescope}, which is operated by the 
Jet Propulsion Laboratory, California Institute of Technology under a contract with NASA.
MHi acknowledges financial support from the Leverhulme trust. This work was supported by the U.S. National 
Science Foundation through awards AST-0408698 and AST-0965625 for the ACT project, as well as awards 
PHY-0855887 and PHY-1214379, along with award AST-0955810 to AJB. Funding was also provided by 
Princeton University, the University of 
Pennsylvania, and a Canada Foundation for Innovation (CFI) award to UBC. ACT operates in the Parque 
Astron\'omico Atacama in northern Chile under the auspices of the Comisi\'on Nacional de Investigaci\'on 
Cient\'ifica y Tecnol\'ogica (CONICYT). Computations were performed on the GPC supercomputer at the SciNet
HPC Consortium. SciNet is funded by the CFI under the auspices of Compute Canada, the Government of Ontario, 
the Ontario Research Fund -- Research Excellence; and the University of Toronto.

\bibliographystyle{mn2e}
\bibliography{refs}

\label{lastpage}

\end{document}